\documentclass{revtex4-1}
\usepackage{graphicx}
\usepackage{epstopdf, epsfig}
\usepackage{hyperref}
\usepackage{subcaption}
\usepackage{epstopdf}

\usepackage{amsmath}
\usepackage{amssymb}

\begin{document}

\title{Simple advecting structures and the edge of chaos in subcritical tokamak plasmas}

\author{Ben F. McMillan}
\affiliation{Centre for Fusion, Space and Astrophysics, Department of Physics, Warwick University, Coventry, United Kingdom}
\author{Chris C. T. Pringle}
\affiliation{Applied Mathematics Research Centre, Coventry University, Coventry, CV1 5FB, United Kingdom}
\author{Bogdan Teaca}
\affiliation{Applied Mathematics Research Centre, Coventry University, Coventry, CV1 5FB, United Kingdom}
%


\begin{abstract}
In tokamak plasmas, sheared flows perpendicular to the driving temperature gradients can strongly stabilize linear modes. While the system is linearly stable, regimes with persistent nonlinear turbulence may develop, i.e. the system is {\it subcritical}. A perturbation with small but finite amplitude may be sufficient to push the plasma into a regime where nonlinear effects are dominant and thus allow sustained turbulence. The minimum threshold for nonlinear instability to be triggered provides a criterion for assessing whether a tokamak is likely to stay in the quiescent (laminar) regime. At the critical amplitude, instead of transitioning to the turbulent regime or decaying to a laminar state, the trajectory will map out the {\it edge of chaos}. Surprisingly, a quasi-traveling-wave solution is found as an attractor on this edge manifold. This simple advecting solution is qualitatively similar to, but simpler than, the avalanche-like bursts seen in earlier turbulent simulations and provides an insight into how turbulence is sustained in subcritical plasma systems. For large flow shearing rate, the system is only convectively unstable, and given a localised initial perturbation, will eventually return to a laminar state at a fixed spatial location.  
\end{abstract}

\maketitle

%
%
%
%
\maketitle

\section{ Introduction} The strong effect of sheared flows on linear plasma instabilities results in a broad range of {\it subcritical} configurations, which are linearly stable but allow long-lived turbulence to develop given a large enough displacement from equilibrium. Such subcritical configurations in plasmas\citep{rincon,friedman15} and tokamaks more specifically\citep{casson, ben_bursts, Roach09, vanwyck} have recently come under extensive study. Computationally, the late-time properties of the turbulent state in a subcritical plasma can be determined by giving the plasma a sufficiently large initial kick\citep{casson}, but whether or not an experimental plasma would enter this regime depends both on the threshold for nonlinear instability, and details of the experimental time-history. 

We specialise to microinstabilities in tokamak plasmas, and use a gyrokinetic model. These equations time-evolve the system state, which is captured by the distribution function $f$, for a set of parameters, which capture the background geometry and plasma profiles. In this article, we investigate the threshold in state space [not parameter space as in some other studies\citep{Highcock}] between the quiescent (laminar) and turbulent state, the {\it edge of chaos}\citep{itano_toh,skufca,chris_bursts}. We examine the dynamics of the plasma on that threshold and well as practical questions about how large an initial perturbation is required to induce long-lived turbulence in a tokamak configuration. The behaviour and solutions found on the edge of chaos potentially provide insight into the domain of existence and nature of the turbulent state. Related questions have been explored via linear theory and dimensional analysis to capture aspects of nonlinear threshold physics\citep{Schek12,Highcock}. Various new tools have been developed to understand the edge of chaos, and applied to neutral fluid theory, and we aim to use these tools to illustrate some questions in plasma physics; an earlier paper\citep{chris_bursts}, studying a drift-wave model (which we refer to as the PI model), discussed the application of these tools and the associated terminology in a plasma context. Essentially, this work can be viewed as a study of whether features found in the {\it edge of chaos} in a simplified drift-wave model\citep{chris_bursts} are qualitatively recapitulated in gyrokinetics. For example, are the relatively simple features of the edge dynamics in the drift wave model, such as the existence of an attractive relative periodic orbit in the edge, a consequence of the simplicity of the drift model, or a robust consequence of the basic physics that will persist even in complicated gyrokinetic scenarios? The complexity and computational intensiveness of the gyrokinetic model compared to the fluid drift model mean that we do not attempt to replicate all the analyses in the earlier paper, and require certain simplifications.

Knowledge of the threshold state potentially provides insight into how nonlinear and linear terms combine to allow quasi-steady states in plasma turbulence.
We find a simple propagating state on the edge of chaos, and this allows us to provide a relatively simple picture of how nonlinear effects can sustain the dynamics in a linearly stable regime. This propagating edge-state mirrors many of the features of the propagating bursts/avalanches\citep{candy_d3d,ben_bursts} seen in the turbulent regime.

Within chaotic motion, simpler exact solutions (steady states, travelling waves, periodic orbits and so on) of the underlying equations are embedded. These solutions are linearly unstable, but their presence can be observed in the dynamics as the flow approaches these states before drifting away. For most subcritical problems, there are two attracting states that the flow can end up in -- laminar flow or statistically steady turbulence. For a given choice of parameters, all initial conditions will evolve into one of these two states depending on which state's basin of attraction the initial condition is in. The only exceptions to this are those states that fall precisely on the boundary separating these two basins. This boundary is referred to as the edge of chaos and any flow which begins on this edge must remain there forever.
 
The edge of chaos can be isolated via an iterative process\citep{chris_bursts}. Although disturbances within the edge must remain within the edge as they evolve, in many systems the dynamics within the edge can still be complex. Typically the flow is chaotic, but the chaos is of a slower, less energetic nature than the fully turbulent flow. As such, there are exact solutions embedded within the edge giving the chaos structure. These structures are linearly unstable, but the number of unstable directions is important. If they only have one, then this direction must be out of the edge and so when the dynamics are restricted to being within the edge, the state becomes stabilised and acts as an attractor \emph{within the edge}  -- that is to say it is an attractor for initial conditions that are in the edge. Such behaviour has been observed in classical shear flows such as pipe flow and plane-Couette flow, but only after sufficient restricting symmetries have been applied.
 
The edge of chaos gives us insight into the full problem in four distinct ways. Firstly, when trying to understand how transition occurs, the edge controls the transition scenario -- to trigger turbulence you must first `cross' the edge, and likewise to relaminarise. Secondly, when seeking to assess how stable the laminar flow is (for instance to assess the effectiveness of control strategies or domain design) the amplitude of the edge is the amplitude required to trigger turbulence -- the larger the amplitude the more stable the laminar state is. Thirdly, the chaos within the edge is simpler and calmer than the full turbulence and so a more ready target for analysis. This analysis should give insight into full turbulence as many of the mechanisms must port across (the equations of motion are the same). Finally, the exact coherent states within the edge which characterise the chaos should be more easily identified than for full turbulence. In all classical flows considered, the exact solutions within the edge can be continued through parameter space to find counterparts of them embedded within full turbulence and hence the associated insight into the full problem.

\section{The system analyzed}
This strongly magnetized plasma system, an idealised tokamak, is described via a gyrokinetic (GK) framework\citep{Hahm_1988}. The GK equations describes particle motion in magnetized plasmas in a self-consistent electromagnetic field by evolving the gyrotropic particle distribution function $f(x,y,\theta,\mu,v_{\parallel})$, where $x,\theta$ and $y$ are spatial coordinates  and $\mu$ and $v_{||}$ are velocity space coordinates. The $x$ coordinate parameterises the radial direction, $\theta$ is the poloidal (straight field line) angle, and the $y$ coordinate is a field line label, with $y \propto (\zeta - q(x) \theta)$, with $\zeta$ the toroidal direction and $q$ the safety factor\citep{fluxtubegeom}. Note that changing $y$ and keeping other spatial variables fixed corresponds to a displacement in the toroidal direction of symmetry. The $x$ and $y$ coordinate are scaled such that $|\nabla x|, |\nabla y| \sim 1$.

\begin{figure}
\includegraphics[width=9cm]{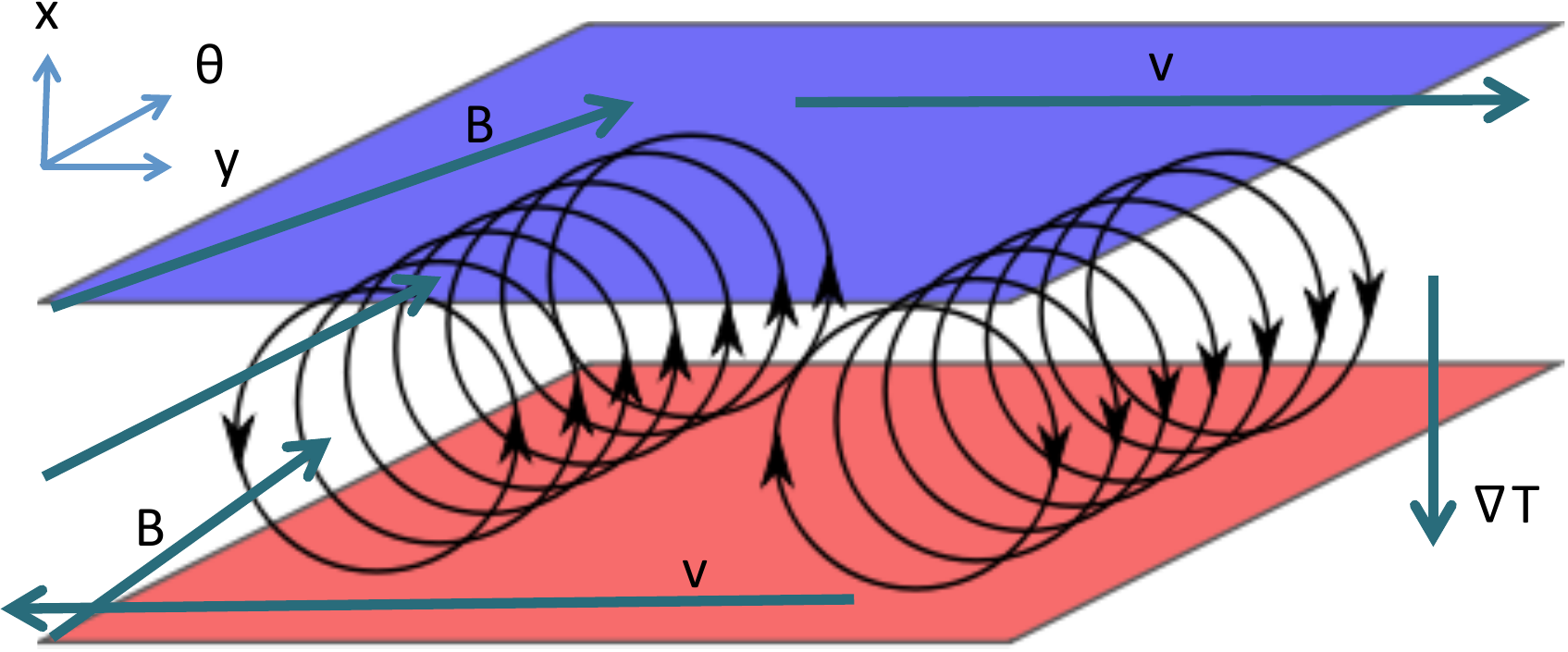}
\caption{Geometry of the plasma system near the outboard mid-plane. Black circles indicate convective vortices generated by the drift instability. The velocity arrows indicate the sheared background flow.}
\label{fig:diag_geom}
\end{figure}

The basic instability driving turbulence in this system is the ion temperature gradient instability, driven by pressure gradients aligned with magnetic curvature. A sketch of the geometry is provided in fig. \ref{fig:diag_geom}. Simulating tokamak turbulence in a kinetic rather than fluid model allows certain details such as parallel Landau damping, perpendicular particle resonances, and finite-Larmor radius effects to be retained. These features are essential to recover a good quantitative match against experimental data: qualitatively, however, many of the dynamical features match a highly simplified 1D model of turbulence\citep{ben_bursts,chris_bursts}. For the GK system of interest, quantities peak near $\theta=0$ and vary slowly along the field line (with field-line spatial dependence resembling linear eigenmodes), and we believe much of the important non-linear dynamics can be understood by examining the $(x,y)$ dependence on the $\theta=0$ {\it mid-plane} in our analysis.

The usual control parameter of interest in plasma microturbulence is the gradient driving the instability, and a normalised measure is the drive rate compared to the parallel streaming time. Due to all spatial scales being subject to Landau damping by parallel streaming, but only some being favourable for instability drive, a somewhat narrow range of wavelengths is strongly activated in a typical simulation (especially in the simplified system here with only one active particle species). Finer scale structures are generated through spatial and velocity-space mixing, but their influence on collective behaviour is reduced due to gyro-averaging and velocity-space integration. Physical or numerical dissipation provides a means of saturation for these fine structures, but quantities like the spectrum of excited modes are often not very sensitive to the value of this dissipation. This should be contrasted with the situation in fluid turbulence where the typical control parameter is the Reynolds number, which directly deterimines the size of the range of wavenumbers that are dynamically relevant. 

Compared to fluids, the overall dissipative nature of kinetic systems is much more complex. 
Collisions\citep{Landau}\citep{Balescu}\citep{Lenard} are ultimately responsible for setting the dissipation scale.
However, in typical under-resolved `collisionless' simulations, anomalous dissipation \citep{Eyink:2018p2095} due to the truncation of the physical nonlinear couplings can be seen as the cause for the
effective removal of energy from fine scales. In addition, hyperviscosity terms, such as the ones employed in our work,
can also play this role. They can be seen as the simplest form of Large Eddy Simulation models, applied to a gyrokinetic system\citep{BanonNavarro:2014p1535}.
Moreover, as the route to dissipation occurs in phase space \citep{Schekochihin:2008p1034}, is linked to the phase space mixing \citep{Tatsuno:2009p1096}\citep{Watanabe2006}  and the energy cascade \citep{BanonNavarro:2011p1274} \citep{Watanabe:2012p1420}\citep{Teaca:2014p1571}, nonintuitive behaviors can emerge \citep{Schekochihin:2016p1902}.
The dissipation for the GK problem occurs over a wide range of spatial scales, including relatively large perpendicular spatial scales \citep{Hatch:2011p1369} \citep{Hatch:2014p1639}, so in conjunction with the injection of energy and subsequent weak energy cascade \citep{Howes:2008p1132}, the overall turbulence dynamics may be considered strongly dissipative. 
For example, in the fluid model of \cite{ben_bursts,chris_bursts}, an explicit dissipation of order 1 in terms of the characteristic scales is applied, which is modelling the effect of kinetic processes;
this dissipative system nonetheless qualitatively captures many of the aspects of the nominally collisionless gyrokinetic system.

\section{Formulation details and symmetry}

We use the GKW code\citep{GKW_paper} to evolve the electrostatic local (flux-tube) gyrokinetic equations, with adiabatic electrons, in the presence of a background poloidal flow shear with shearing rate $S$. The aim is to focus on a somewhat simplified system, in which analysis is relatively straightforward, rather than to perform a detailed device-relevant full-physics model. Our analysis treats this fairly standard simulation setup in some ways as a `black-box', so much of the detail presented in this section will not be referred to in later discussion. Despite this, we present the equations of this GK system for completeness, in direct space (rather than spectral) form; more details are provided in the code reference paper\citep{GKW_paper}. The dynamics are found by solving a Vlasov equation for the perturbation $\delta f$ to the distribution function,
\begin{equation}
 \frac{ \partial \delta f }{ \partial t} = - \dot{Z}_0 \cdot \nabla_Z \delta f - \dot{R}_1 \cdot \nabla_R (f_0 + \delta f ) - \dot{v}_{||1} \frac{\partial}{\partial v_{||}} f_0  + C(\delta f)
\end{equation}
where $f_0$ is the (Maxwellian) background distribution function, $Z$ represents the 5D phase space $(x,y,\theta,\mu,v_{\parallel})$, $R$ the spatial coordinates $(x,y,\theta)$, $\dot{Z}_0$ are the drift trajectories in the absence of the perturbing electrostatic field, $\dot{R}_1$ is the $E \times B$ drift, and $\dot{v}_{||1}$ represents accelaration due to $E_{\parallel}$. $C$ stands in for the collision operator(which is not used here), or for the numerical hyperviscosity.

For the zero-$\beta$, weak-flow, electrostatic case of interest, for ions of charge $e$, and equal ion and electron temperatures,
the equations of motion may be written as
\begin{equation}
  \dot{Z}_0 = (\dot{R}_0, \dot{v}_{||0}, \dot{\mu_0}) = (v_{||} \mathbf{b} + \mathbf{v}_D + \mathbf{v}_{E0}, -\frac{\mu B}{m} \frac{\mathbf{B} . \nabla {B}}{B^2}, 0) 
\end{equation}
\begin{equation}
  \dot{R}_1 =  \mathbf{v}_E
\end{equation}
\begin{equation}
  \dot{v}_{||1} = \frac{ 1 }{m v_{||} } ( -e [ v_{||} \mathbf{b} + \mathbf{v}_D] . \nabla \langle \phi \rangle_\alpha - \mu \mathbf{v}_E . \nabla B )
\end{equation}
with
\begin{equation}
  \mathbf{v}_E = \frac{\mathbf{b} \times \nabla \langle \phi \rangle_\alpha}{B}
\end{equation}
where angle brackets with subscript $\alpha$ denote a gyroaverage, $ \mathbf{v}_{E0} =  \mathbf{E}_0 \times \mathbf{b}/B $ and
\begin{equation}
  \mathbf{v}_D = \frac{1}{e} \left[ \frac{m v_{||}^2}{B} + \mu \right] \frac{\mathbf{B} \times \nabla B}{B^2}.
\end{equation}

For the derivatives of the background distribution function we use
\begin{equation}
  \frac{\partial f_0}{\partial v_{||}} = f_0 \frac{m v_{||}}{ T }
\end{equation}
and 
\begin{equation}
  \nabla f_0 = - f_0 \left( \frac{m v_{||}^2 + \mu B -3 T/2 }{T} \frac{1}{L_T} + \frac{1}{L_n} \right) \nabla x
\end{equation}
with
\begin{equation}
    f_0 = \frac{n}{( 2 \pi T)^{3/2}} \exp{ -\frac{m v_{||}^2 + \mu B}{T} }
\end{equation}
where $L_n$ and $L_T$ are the density and temperature gradient scale lengths. The local limit allows taking $n$, $T$,
$L_n$ and $L_T$ constant. The background electric field $\mathbf{E}_0 = x B_0 S \nabla x$ so that the shearing rate $ d (\nabla y . \mathbf{v_{E0}} )/d x \sim S$.

The quasineutrality equation relates the gyroaveraged charge density associated with $f$ to the perturbed electric field $\phi$,
\begin{equation}
  \int dZ \frac{e}{T} f_0 \left< \left<\phi \right>_{\alpha} - \phi \right>_{\alpha} - \frac{ne}{T} \left( \phi - \langle \phi \rangle \right) + \int dZ e \langle \delta f \rangle_{\alpha} = 0. 
\end{equation}
In the long wavelength limit, explicit calculation of the first term on the right hand side yields $(m n/e B^2) \nabla^2 \phi $, which is the charge associated with the polarisation response of the ions. The second term represents the adiabatic electron response, with the angle bracket (without a subscript) indicating a zonal average
(volume-weighted integration in $y$ and $\theta$ direction) as the electrons are bound to the flux surfaces and do not respond to zonal charge fluctuations. 

The symmetries of these gyrokinetic equations may be found by inspecting the form of these terms.
The electrostatic field and hence the $E \times B$ drift can be expressed as a linear function of $\delta f$, so we have $ \dot{Z}_1 = L ( \delta f )$. Due to axisymmetry and the flux-tube limit, both $L$ and $\dot{Z}_0$ are spatially invariant to translations in the $x$ and $y$ directions. The boundary conditions in the $x$ and $y$ directions are doubly periodic, but in the $\theta$ direction, there is a twisted periodicity of the form
$f(x,y,\pi) = f(x,y + s x,-\pi)$, with $s$ dependent on magnetic shear. As a consequence, the overall system has a continuous symmetry in the $y$ (toroidal) direction, but only a discrete (not continuous) translation symmetry in the $x$ (radial) direction.
There is also an inversion symmetry\citep{ParraBarnes2011}, which involves changing the sign of one of the parameters (the flow shear), and allows us, as usual, to consider only cases with $S \ge 0$, since results for $S < 0$ may be found using the symmetry. For example, propagating structures exist with the same radial velocity but oppisite sign for $S < 0$.

In the following, units for amplitudes use the local gyrokinetic convention, so the electrostatic potential $\phi$ is in units of $\phi_0 = \rho^* e \phi / T_e$, with $\rho^* = \rho_0 /R$, so fluctuations, although order $1$ in these plots, are small in terms of relative density (here, $T_e$ is the electron temperature $\rho_0 = (m T_e)^{1/2} / q B$ is the `ion sound gyroradius', $R$ is the tokamak major radius). 

The simulations use the standard set of CYCLONE parameters, with $L_n = R/2.2$, $q=1.4$, $(dq/dr) r/q = 0.8$, but with a slightly reduced temperature gradient, $L_T = R/6.0$, and a concentric circular equilibrium, with local aspect ratio $0.18$. The size of the simulation box is $[L_x,L_y] = [157, 84] \rho_0$, with 20 toroidal modes used, $321 x$ grid points, and $16, 16$ and $32$ grid points used in the $\theta$, $\mu$ and $v_{||}$ directions. A normalised forth-order numerical hyperviscosity parameter of $0.1$ is chosen in the parallel, $v_{||}$ and $x$ directions (in addition to inevitable numerical diffusion); this corresponds to damping oscillations at the grid-scales in these directions on a timescale of $10 t_0$, and helps to avoid numerical problems of spectral pile-up at fine-scales without unduly influencing the longer scales that will be of interest here.

Internally, GKW represents the distribution function using a Fourier series. Except where specified otherwise, the simulations use an initialisation of the form
\begin{equation}
  f(k_x,k_y,\theta,v,\mu) = A f_0(v,\mu) \exp\left\{ -\frac{(k_x-k_{x0})^2}{ C_x^2} -\frac{(k_y-k_{y0})^2}{ C_y^2} \right\} \frac{1}{2} \left( \cos(\theta) + 1 \right)
  \label{eq:initialise_wavepacket}
\end{equation}
representing a field-aligned density fluctuation with typical wavenumber $(k_{y0}, k_{x0})$ modulated by a gaussian envelope in the $x$ and $y$ directions with width parameterised by $C_x$ and $C_y$, with an overall amplitude $A$. Unless noted otherwise, values $k_{x0} \rho_0 = 0.24$,  $C_x \rho_0 = 0.1601$, $k_{y0} \rho_0 = 0.37$, $C_y \rho_0 = 0.074$ will be used.

\section{The edge state}
\begin{figure}
\includegraphics[width=9cm]{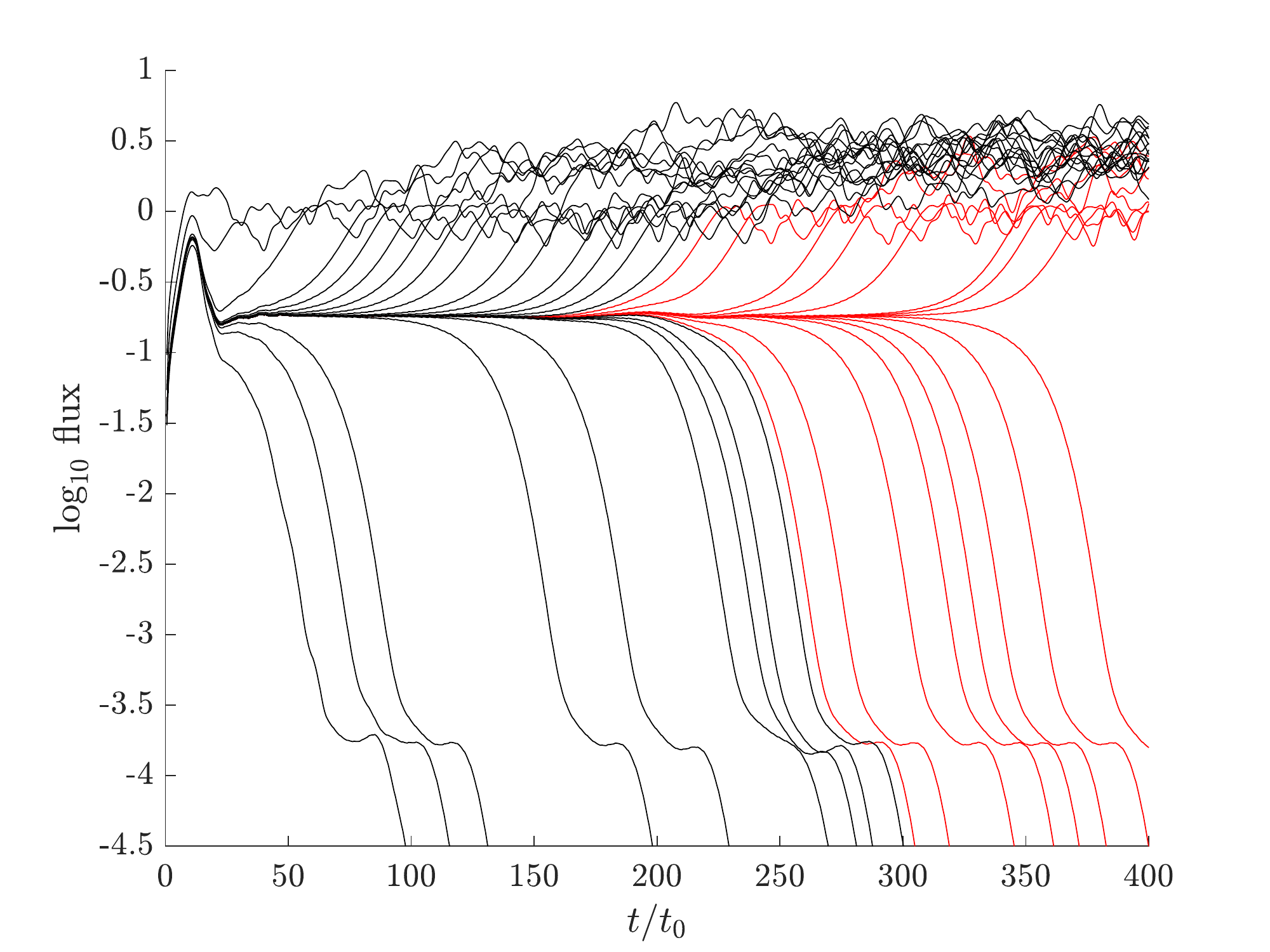}
\caption{ Heat flux versus time (gyrobohm units) for simulations with $S = 0.12 t_0^{-1}$ and successive initial condition
  amplitudes chosen using a bisection method to approach the critical amplitude. Red traces are restarted from $t=120$, with the distribution function rescaled to track the edge state.}
\label{fig:initscan}
\end{figure}

The black traces of Fig.\ref{fig:initscan} show the evolution of the heat flux in the system for initializations of varying amplitude versus time (in units of transit frequency $t_0 = c_s/R$). The linear system is stable despite periods where the flux increases in time, and simulations with sufficiently small amplitude initializations decay. Given a large enough initialization, however, sustained turbulence is triggered. 

The amplitude of the initial perturbation was systematically varied, by a bisection technique\citep{chris_bursts}, to find the threshold amplitude below which the system decays to a laminar state, but above which it remains in a turbulent regime at late time. The simulations very close to threshold remain for some time near the separator between the stable and unstable manifold in the system, i.e. the {\it edge of chaos}, before diverging away. In Fig.\ref{fig:initscan} the near-stationary flux ($\log_{10}$[flux] $ \approx -1$) of the edge state indicates that the edge dynamics are considerably simpler than the turbulent dynamics. The `steps' that appear in decaying simulations ($\log_{10}$[flux] $\approx -4$) are associated with a time dependent (`Floquet mode') behaviour\citep{WaltzTrans}; in this case these dynamics are too slow to play a role in the transition to turbulence.

The edge of chaos represents the separatrix between the attractors for the laminar and turbulent dynamical
states, and is an unstable manifold for the system. When the dynamics are restricted to the edge (by careful choice of initial trajectory),
however, we find a local attractor within the edge, which we refer to as the edge state.
To analyze this state, in addition to standard simulations, we use a series with a very small y domain (narrow simulations), one-fifth the size
of the standard domain (in units of the thermal gyroradius $\rho_0$). In the narrow simulations, the non-zonal component is dominated by the longest-wavelength mode that fits in the
y direction (at late time more than 90\% of the vorticity is in this mode). We use the narrow simulations to focus more directly
on the relevant dynamics in a simpler system with fewer degrees of freedom. The properties of the edge state are qualitatively and quantitatively very similar for standard and narrow simulations.

We also considered simulations initialised from a {\it white noise} perturbation, where independent normally distributed pseudorandom numbers are loaded into the numerical grid of the distribution function as an initial condition, and multiplied by the background distribution function $f_0$. Performing the same bisection method to search for the critical amplitude yields an effectively identical edge state (shifted in the $x$ direction). The insensitivity of the result to the initial perturbation is an indication that the edge state is in fact an attractor within the entire edge manifold. 



\begin{figure*}
\begin{subfigure}{0.58\textwidth}
  \includegraphics[width=8cm]{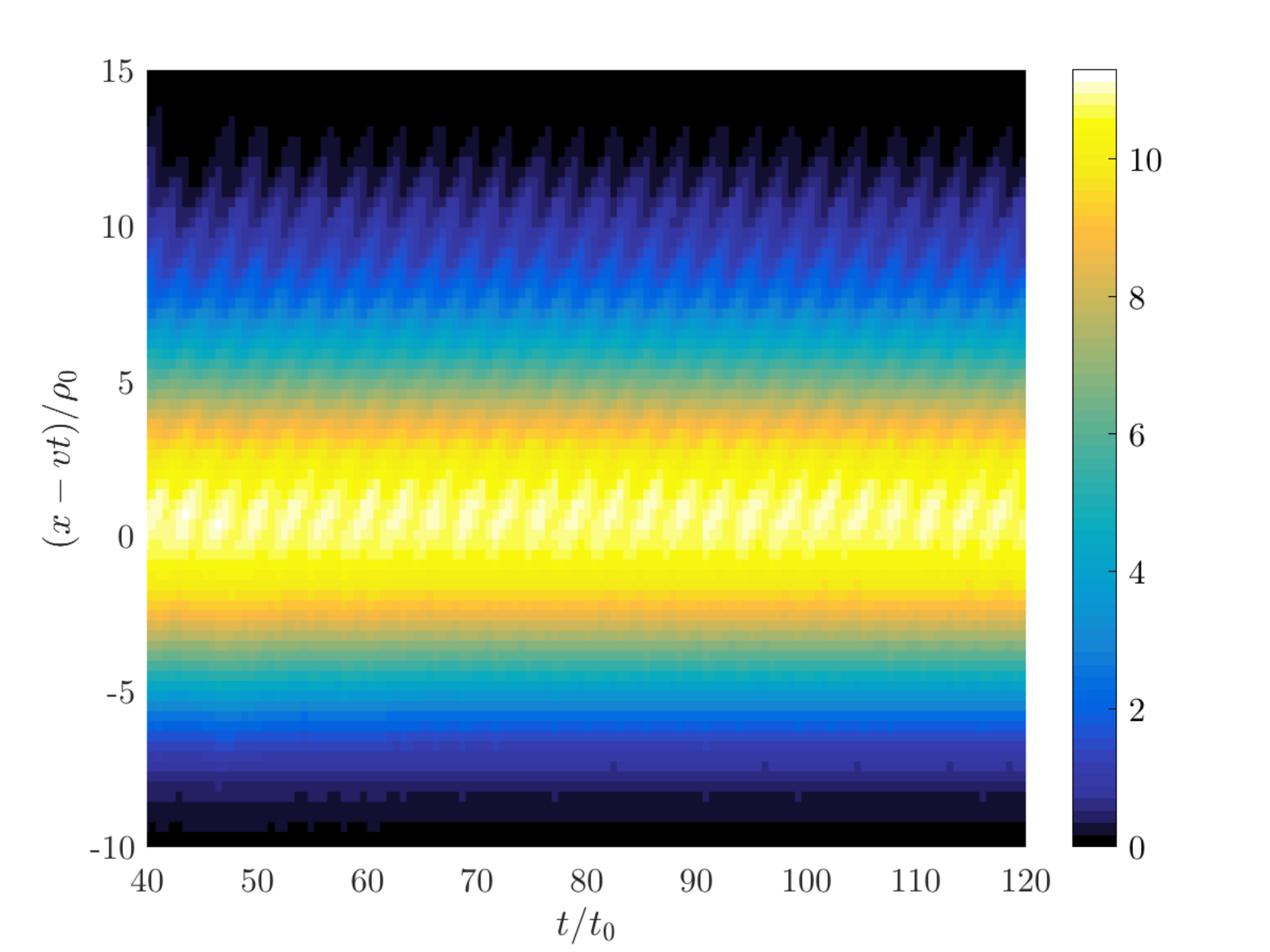}  \caption{} \label{fig:burst_phi2}
\end{subfigure}
\begin{subfigure}{0.38\textwidth}
  \includegraphics[width=4.5cm]{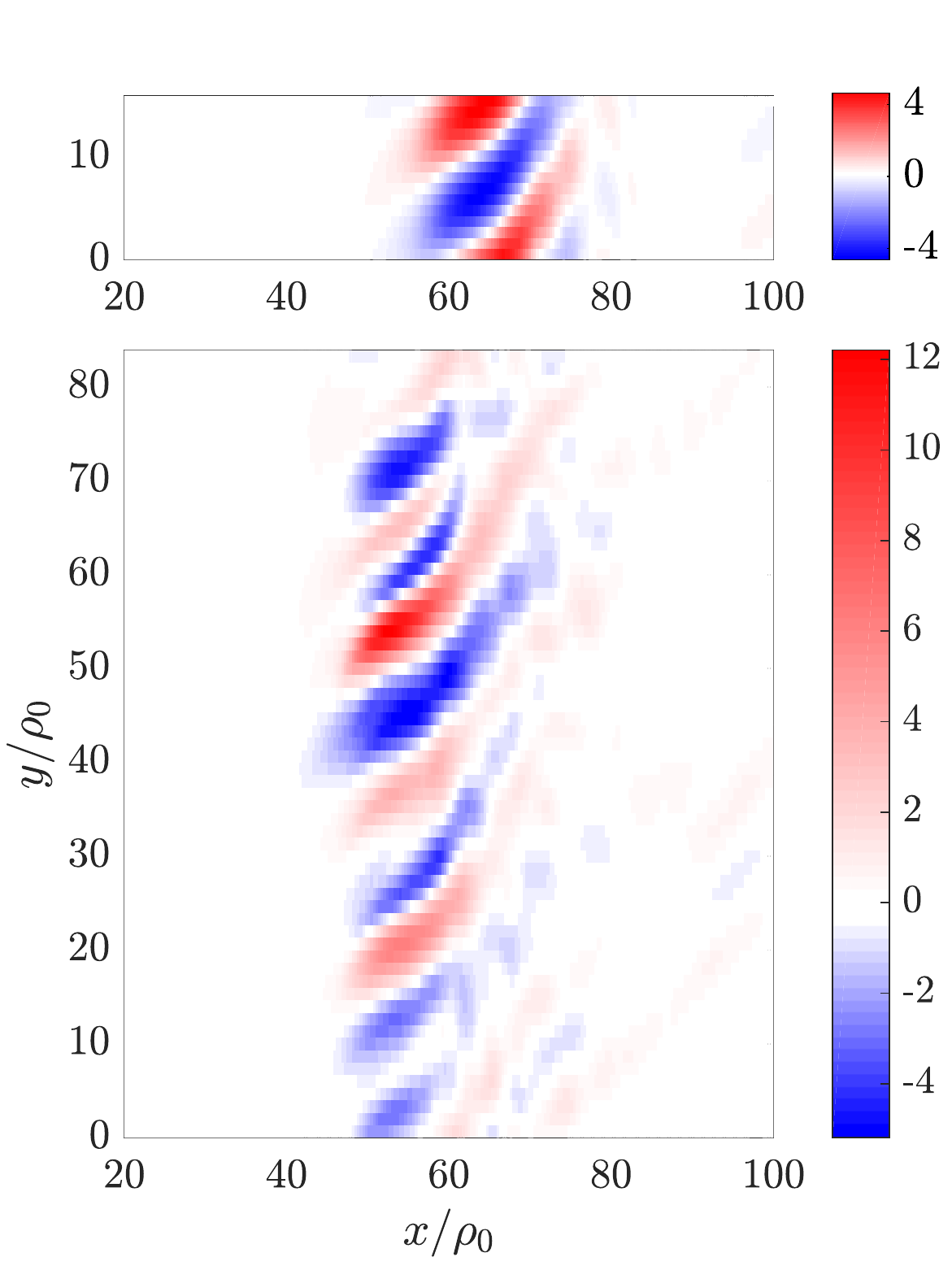}  \caption{} \label{fig:edgestatesnapshot}
\end{subfigure}
\caption{(a) Mean $\phi^2$ (averaged over $y$) at the midplane versus time and position in the
  travelling wave frame $x - vt$ for the edge state in narrow simulations with $S = 0.12 t_0^{-1}$. A periodicity over $3.2 t_0$ is visible.
(b) Non-zonal potential $\phi$ at outboard mid-plane versus $x$ and $y$ for edge state at $t = 120 t_0$ for $S = 0.12 t_0^{-1} $ for (top) narrow and (bottom) standard simulations. }
\label{fig:burst_phi_combined}
\end{figure*}

For narrow simulations, the edge state is found to be very close to a traveling wave. We determined the radial velocity $v$ of this translating structure using a linear fit of the $x$ position of the peak RMS amplitude of the non-zonal potential $\phi^2$ versus time. Detailed inspection (Fig.\ref{fig:burst_phi2}), reveals a small time oscillation, with period ($3.2 t_0$) equal to the distance between lowest order rational surfaces in the system (here there are 60 of these surfaces in the domain) divided by the traveling wave velocity. This is a consequence of the fact that for finite magnetic shear, local gyrokinetics has a discrete, rather than continuous, spatial translation symmetry. The edge state for narrow simulations is thus a relative periodic orbit rather than an exact traveling wave. The RMS variation from exact periodicity (when sampled once every $3.2 t_0$) is found to be $0.5$\% over a period $64 t_0$. For the standard simulations, plots of zonal quantities solutions also suggest a near-periodic orbits on the 12-fold discrete radial symmetry of the larger system, with zonal averages similar to the narrow simulations.


\begin{figure}
\begin{subfigure}{0.5\textwidth}
  \includegraphics[width=7cm]{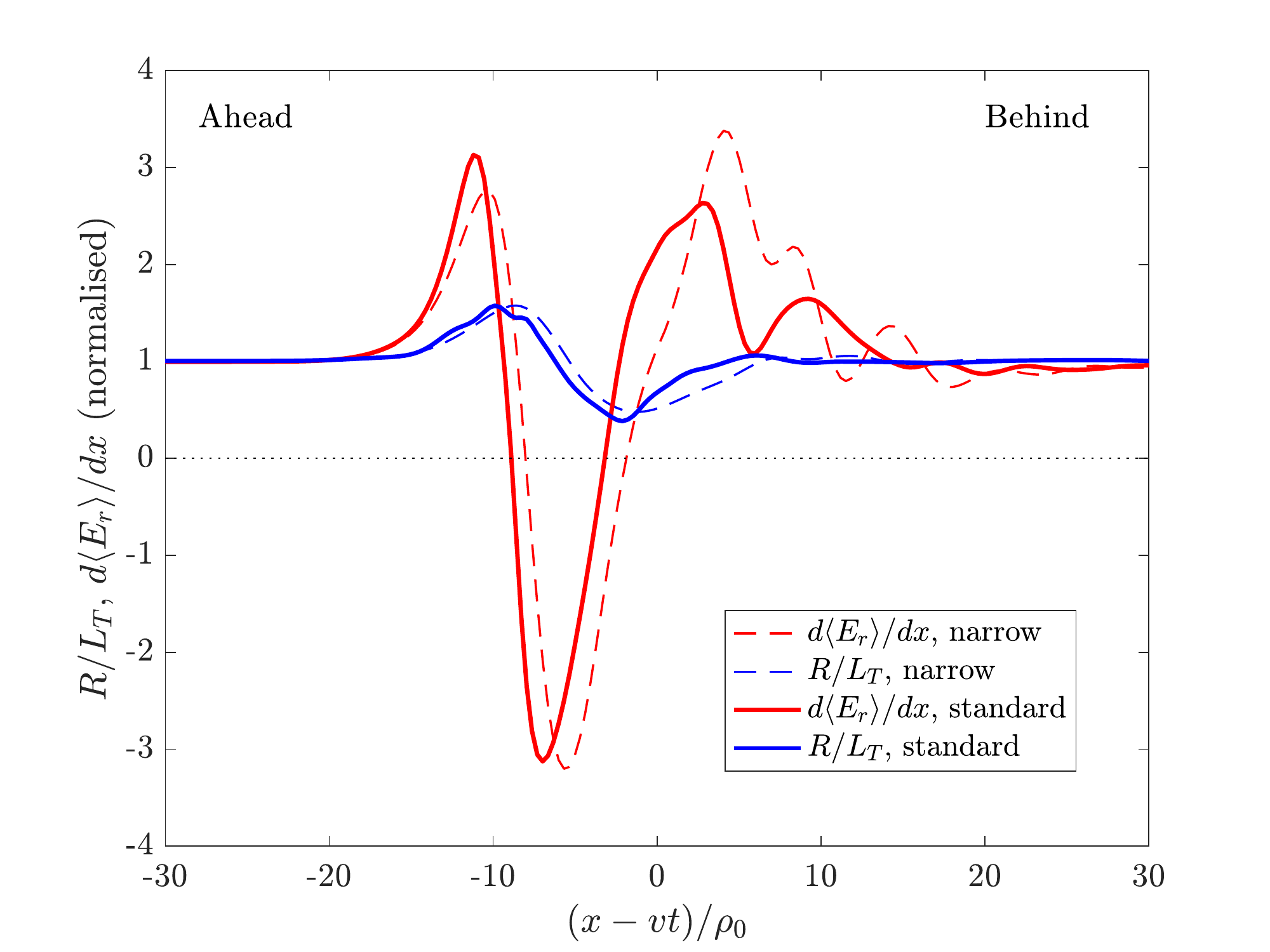}\caption{}
  \label{fig:tempzonal_edge}
\end{subfigure}
\begin{subfigure}{0.5\textwidth}
  \includegraphics[width=7cm]{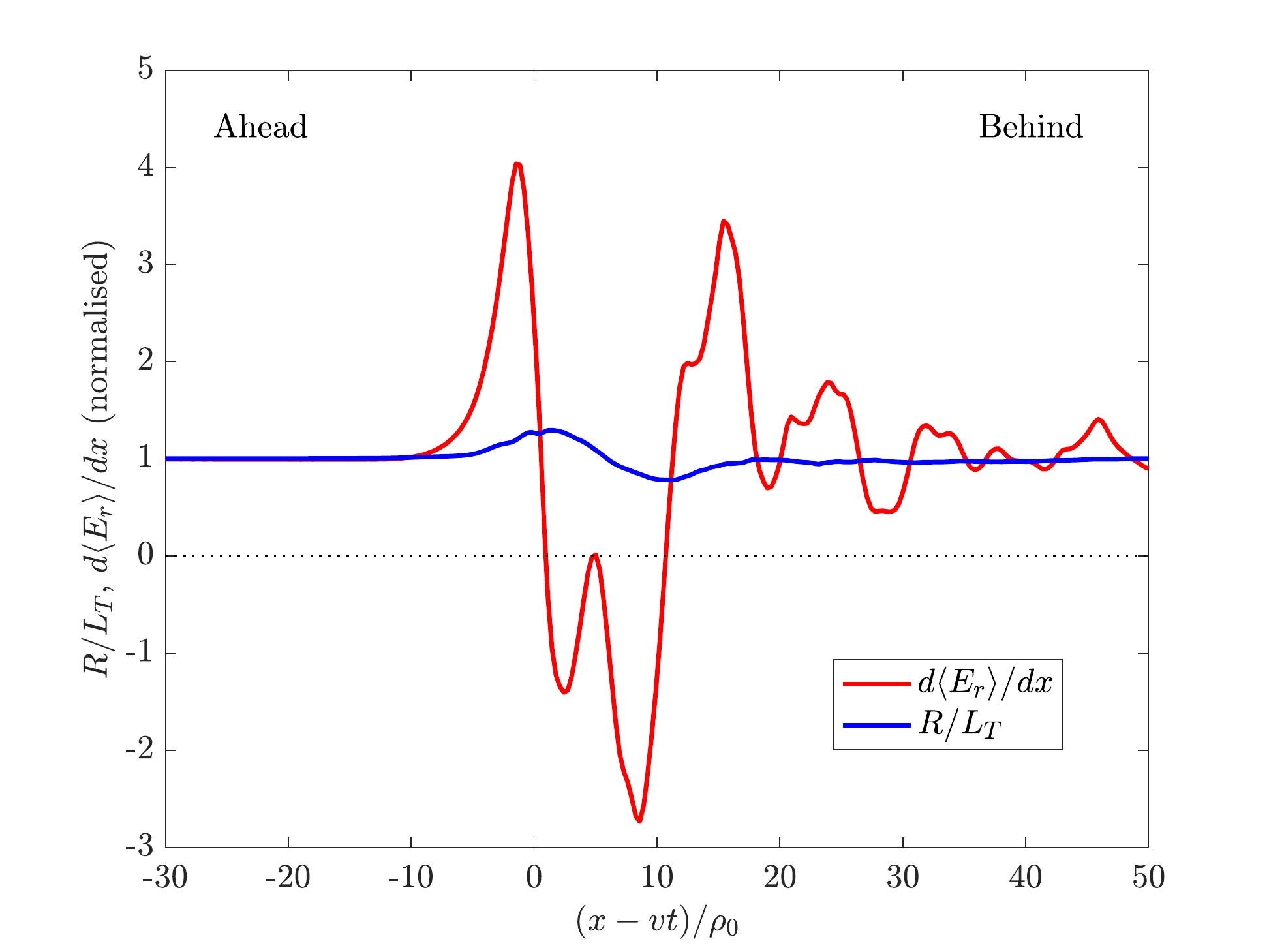}\caption{}
  \label{fig:tempzonal_turb}
\end{subfigure}
\caption{ (a) Mean (time-averaged from $40 - 120 t_0$) temperature gradient (blue) and zonal shear flow $d\langle E \rangle/dr$ (red), both normalized to background gradients, versus position $x-v t$ (in the traveling wave frame) of the edge state, for $S=0.12$, and both narrow (dashed) and standard (solid) simulations.
          (b) Mean (time-averaged from $90 - 150 t_0$) temperature gradient (blue) and zonal shear flow $d\langle E \rangle/dr$ (red), both normalized to background gradients, versus position $x-v t$ (in the approximate burst frame) for a turbulent simulation state, for $S=0.16$.}
\label{fig:tempzonal}
\end{figure}

\begin{figure*}
\begin{subfigure}{0.32\textwidth}
\includegraphics[width=4.6cm]{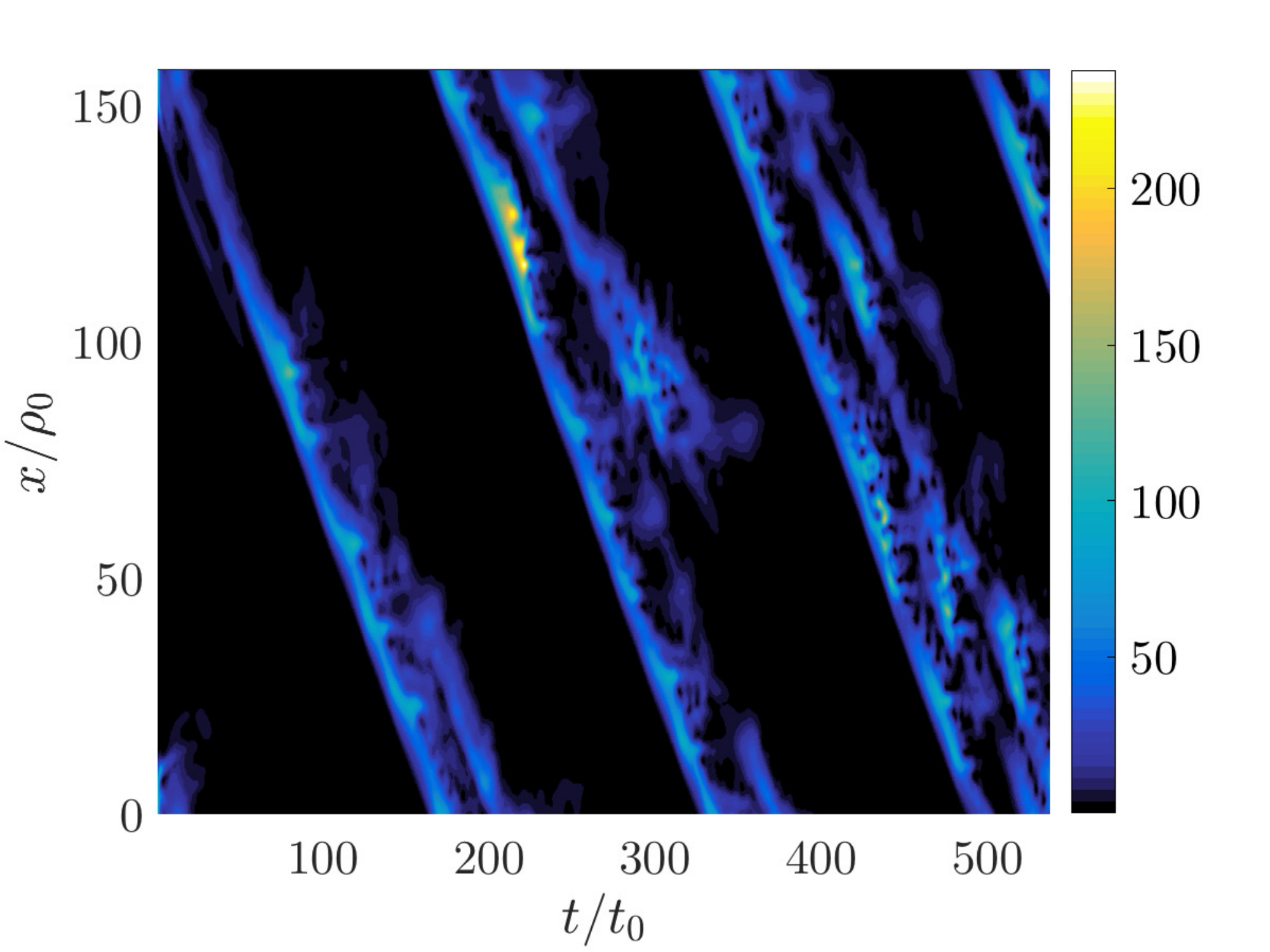} \caption{} \label{fig:timeevol_sp15}
\end{subfigure}
\begin{subfigure}{0.32\textwidth}
\includegraphics[width=4.6cm]{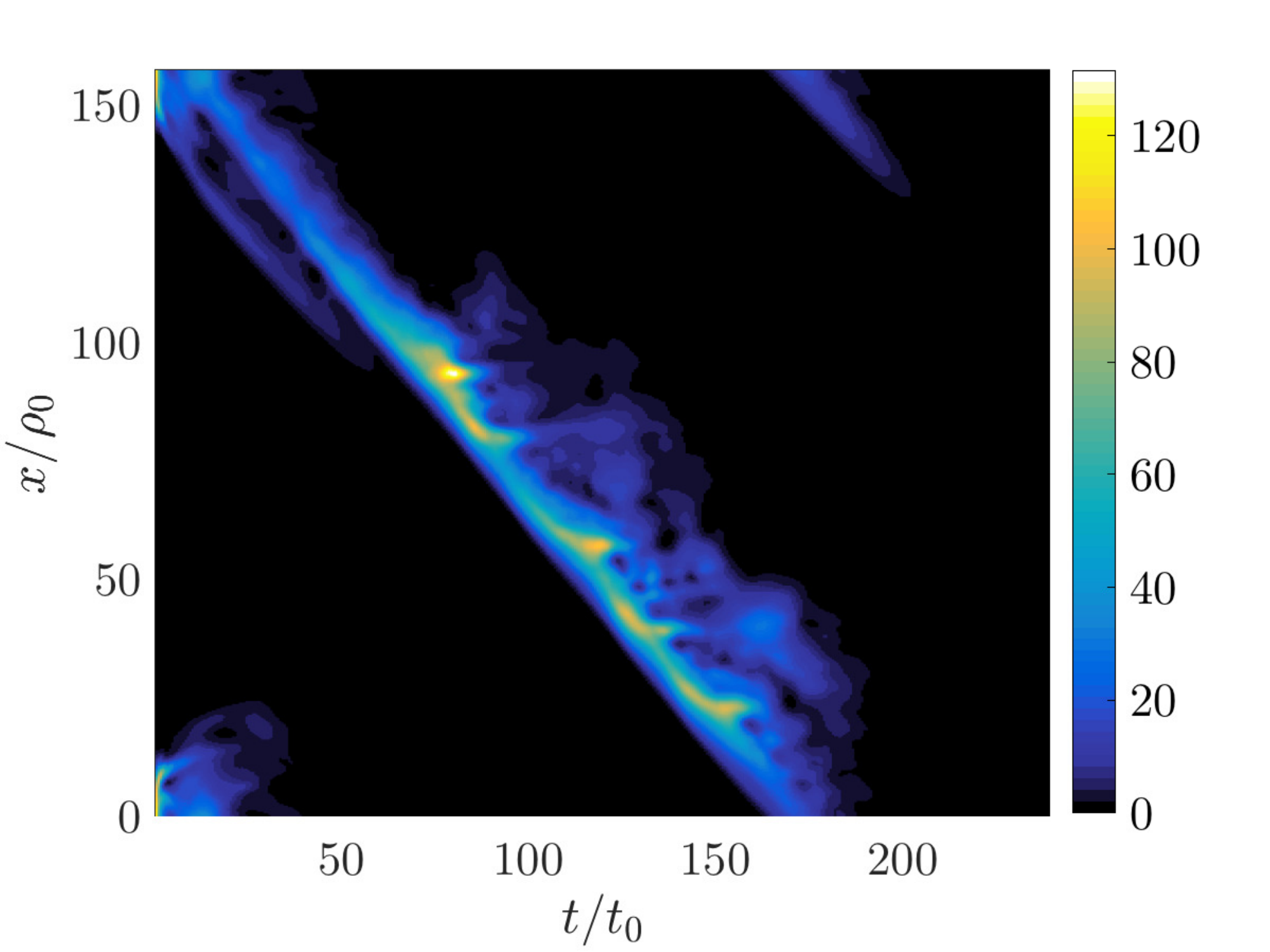} \caption{} \label{fig:timeevol_sp16}
\end{subfigure}
\begin{subfigure}{0.32\textwidth}
\includegraphics[width=4.6cm]{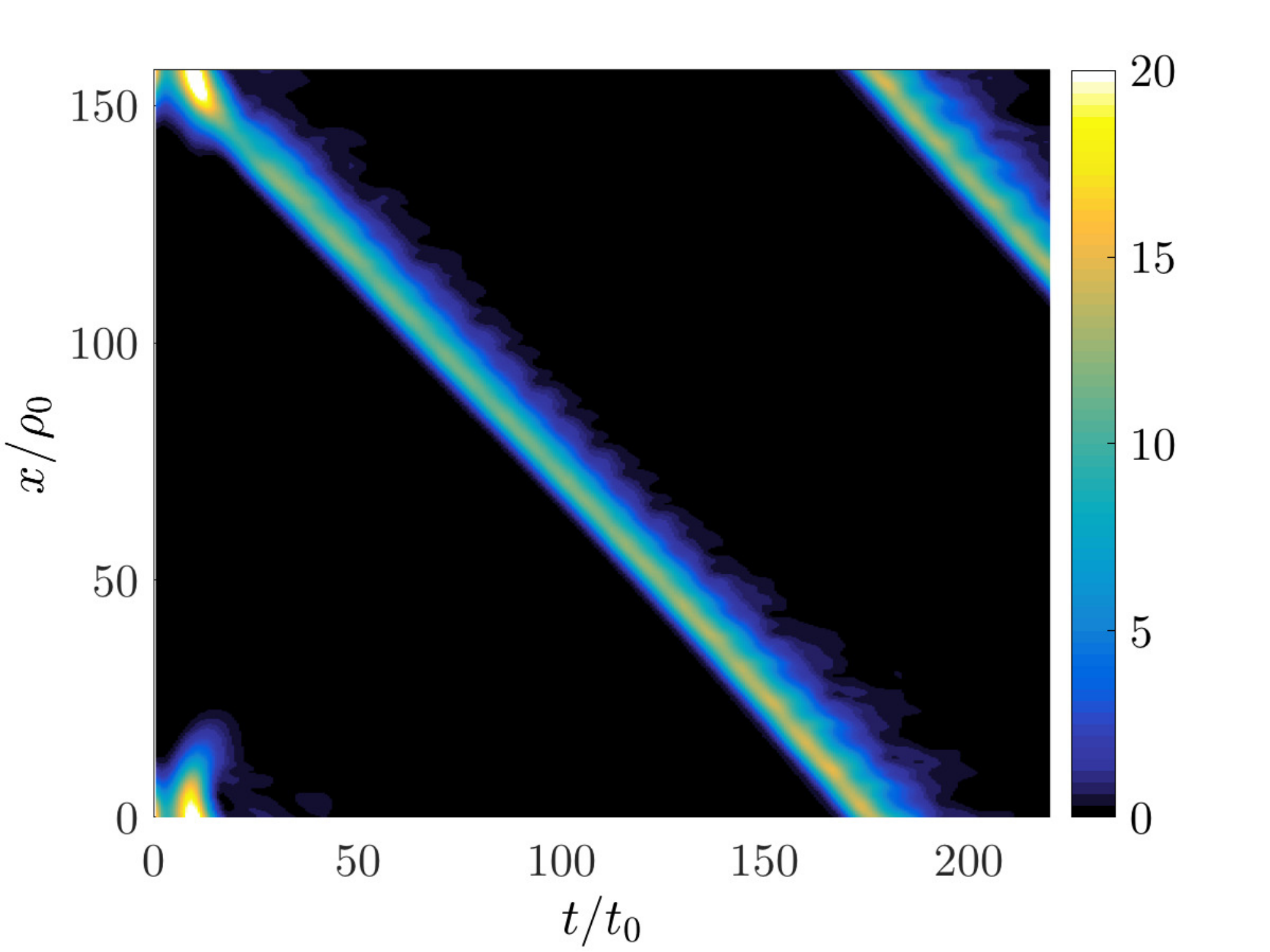} \caption{} \label{fig:phiamp_edgestate_sp12}
\end{subfigure}
\caption{Mean of squared non-zonal potential $\phi^2$ (averaged over $y$) at the mid-plane versus time and $x$ for (a) $S = 0.15 t_0^{-1} $, (b) $0.16 t_0^{-1}$, and (c) for an edge state with $S = 0.12  t_0^{-1}$. In (a), long-lived turbulence is seen in a slowly expanding region centered around the excitation front. In (b), turbulence is excited transiently over a period of $100 t_0$, remaining localized near the traveling excitation front but then decays. The edge state (c) is much simpler and smoother, but also of considerably lower amplitude.}
\label{fig:timeevol_15_16}
\end{figure*}

The quasi-traveling wave solution in both narrow and standard simulations (Fig.\ref{fig:edgestatesnapshot}) consists of a tilted finite $k_y$ traveling wave mode, fed by the gradient-drive, that produces a traveling zonal shear flow ahead (leftwards) of the pulse, that opposes the background shear flow (Fig. \ref{fig:tempzonal_edge}). The traveling perturbation strengthens the temperature gradient ahead of the pulse, and weakens the gradient behind, as expected from the energy transport equation, given the localized heat flux associated with the burst (the change in gradient in fig. \ref{fig:tempzonal_edge} is of comparable size to the background gradient). Those two mechanisms would be compatible with a traveling wave in either direction\citep{ben_bursts,chris_bursts}, but when the nonlinearity in the simulation is turned off, the $k_y\neq 0$ mode amplitude continues to propagate (not shown) in the same direction for $10 a/c_s$ due to the group velocity, which depends on the mean $k_x$ value and thus the sign of $S$\citep{ben_bursts}. Note that narrow simulations do not permit a vortex pair\citep{Has_Mini_inverse} advection mechanism, where a spatially localised `blob' self-advects across the domain, as they are dominated by one $k_y$ mode (unlike $y$-localised features seen elsewhere\citep{Wyk17}). Time snapshots (fig \ref{fig:edgestatesnapshot}) of the mid-plane potential for narrow and standard simulations show similar tilting and localization in $x$ for the two simulations, but despite close similarities in $y$-averaged diagnostics, the standard simulation does not decay towards the narrow edge state. The spatial scale of the edge structure is of order $10 \rho_0$ in the $x$ direction, and the typical wavelength in the $y$ direction is approximately $16 \rho_0$; this is of similar scale to the wavelength of the most unstable mode, at $k \rho_0 \sim 0.5$; this is also comparable to typical scales in the fully turbulent simulations here and elsewhere\citep{CYCLONE}. The combination of linear physics and nonlinear interaction with zonal flows that set the relevant length scales in the turbulence physics\citep{Plunk2015} also appears to be responsible for sustaining the edge state. So the radial ($x$ direction) scales of the edge state might also be estimated by considering the wavenumber of the secondary instability that drives zonal flows\citep{RogersPRL}; this also gives a scale in the $x$ direction of roughly $10\rho_i$. 

\begin{figure}
\includegraphics[width=9cm]{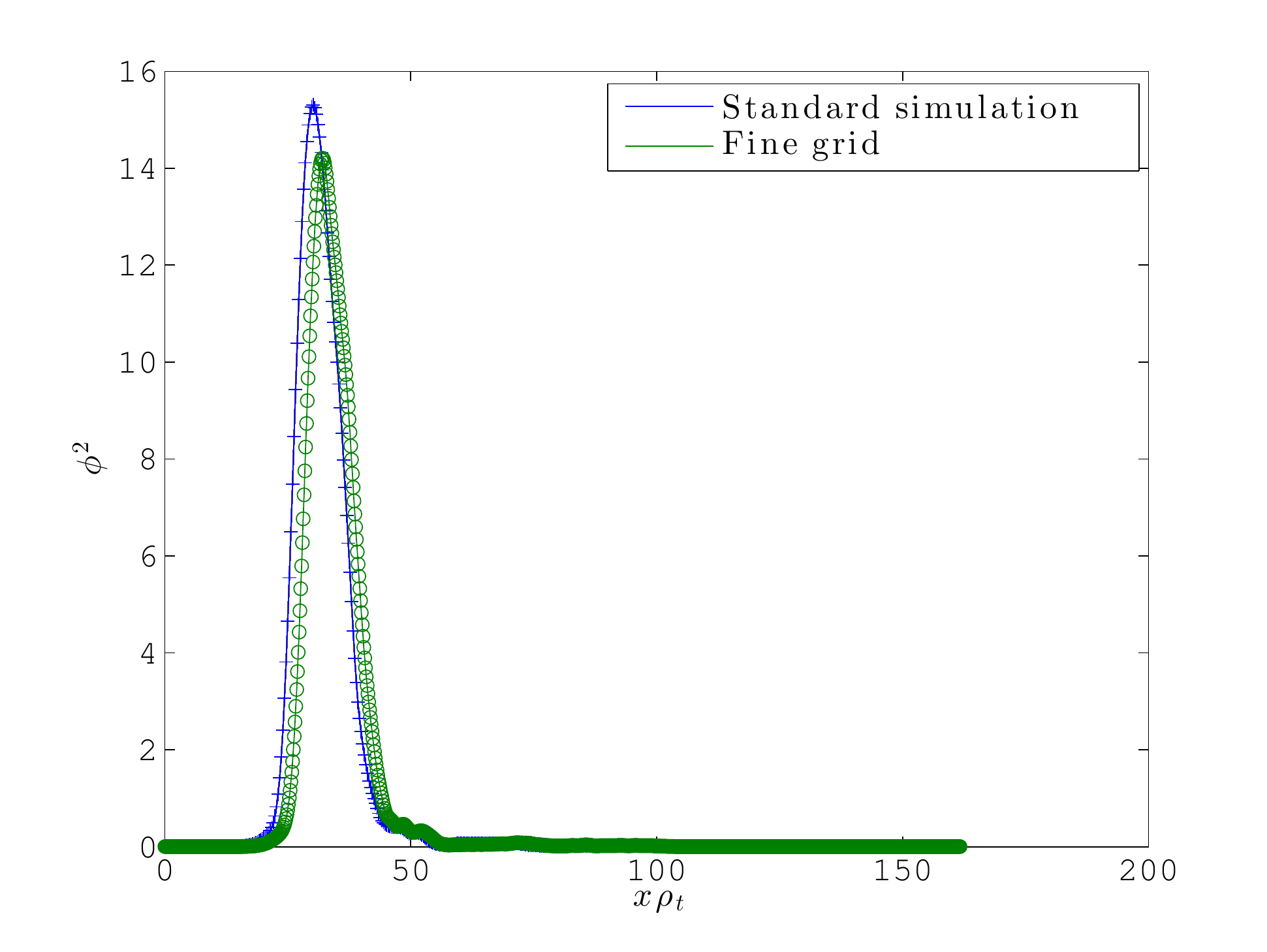}
\caption{Intensity of the non-zonal field $\phi$ (averaged over the $y$-direction) versus $x$ for $t=60 t_0$ at the mid-plane for the edge-state found in a standard (blue trace) and doubled-resolution simulation (green trace).}
\label{fig:refcomp}
\end{figure}

Even though the numerical resolution chosen is known to be sufficient to obtain converged results for turbulence simulations of this case, it is possible that the slightly different quantities of interest related to the edge state require higher numerical resolution. We therefore found the edge state (using the bisection method) in simulations with doubled resolution in each of the five phase space directions, with the hyper-diffusivity in code units reduced by a factor of $16$, for $S=0.12$. Qualitatively, there are no striking differences observed (fig. \ref{fig:refcomp}); the propagating edge state obtained for the high resolution simulation has a mean squared amplitude, width, and propagation velocity within $9\%$, $2\%$, and $1\%$ of that found in standard simulations (with these quantities averged over the time period $45-75 t_0$). We therefore conclude that these phenomena are very insensitive to the value of the numerical diffusivity in the simulation and numerical resolution in general.

\section{Transition to a turbulent state}
\begin{figure*}
\begin{subfigure}{0.49\textwidth}
\includegraphics[width=5.9cm]{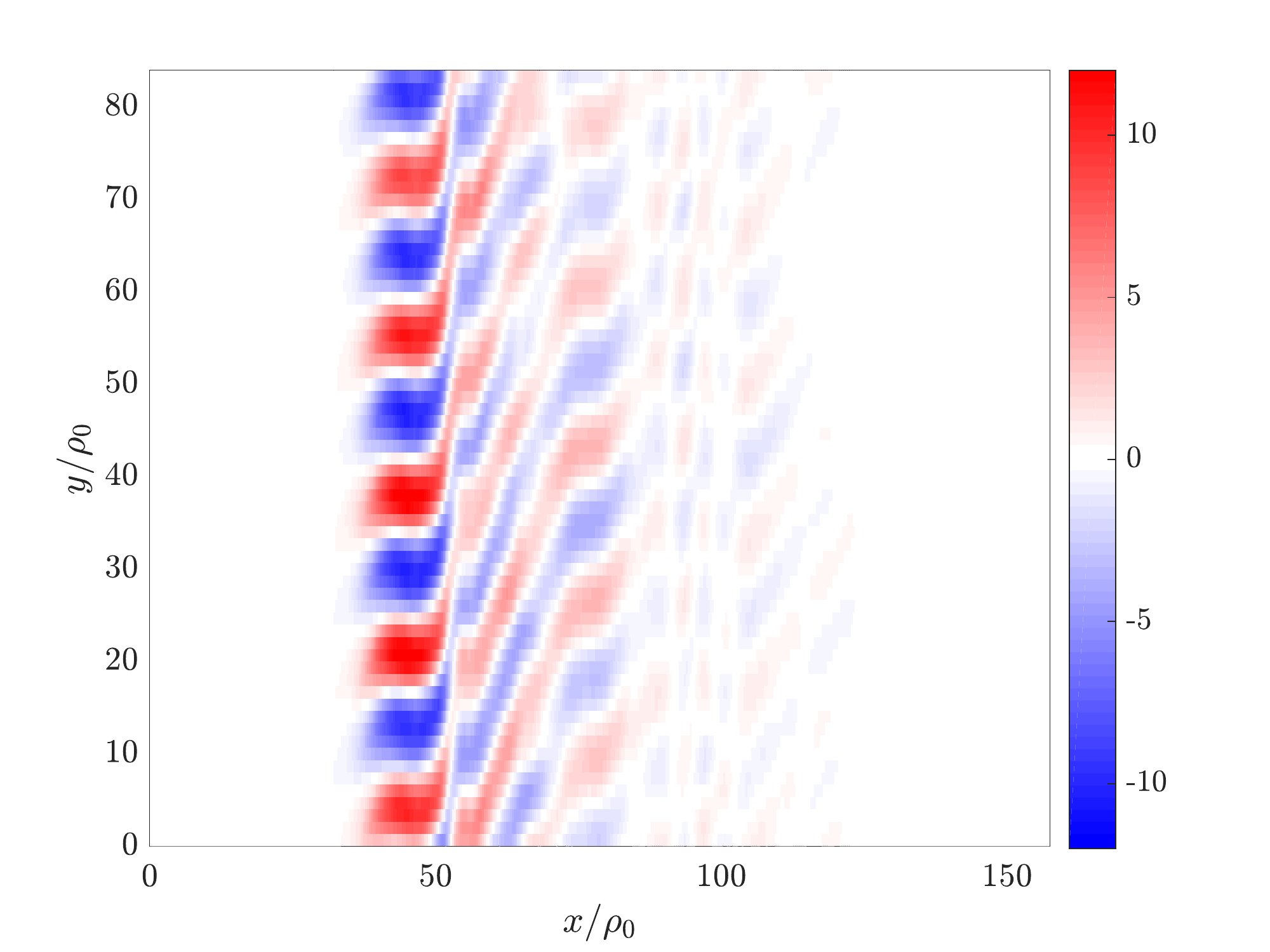} \caption{} \label{fig:critstate_standard}
\end{subfigure}
\begin{subfigure}{0.49\textwidth}
\includegraphics[width=5.9cm]{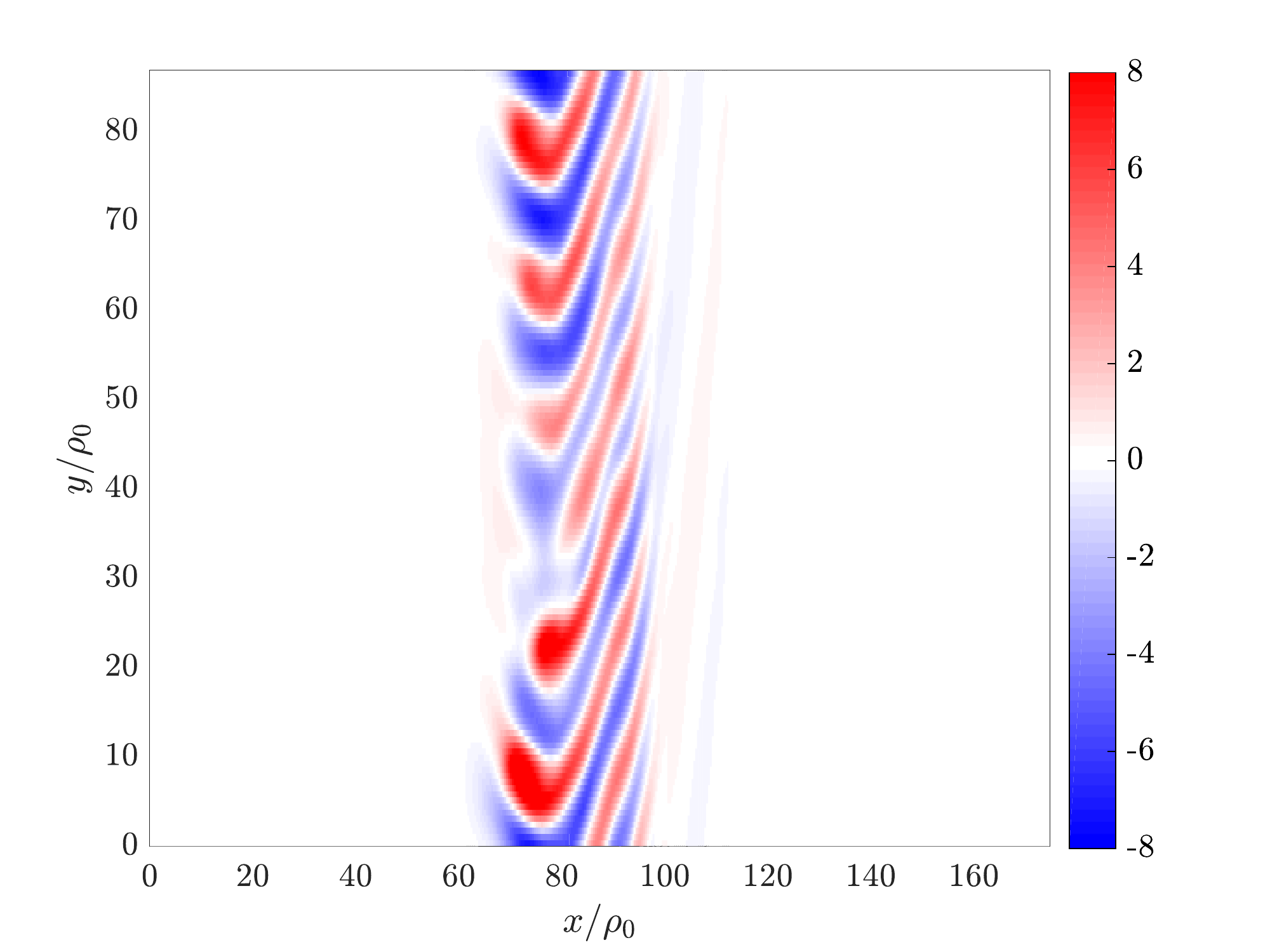} \caption{} \label{fig:critstate_lowshear}
\end{subfigure}
\caption{ Non-zonal potential $\phi$ at outboard mid-plane versus $x$ and $y$ for (a) the standard simulation at near-critical shear value $0.15$ and (b)
          a low-aspect ratio simulation with zero magnetic shear at a near-critical shear value.}
\label{fig:critstate}
\end{figure*}

In typical simulations with a uniform shear flow, an isolated perturbation of sufficient amplitude produces a spreading region of turbulence. For sufficient shear flow (Fig.\ref{fig:timeevol_15_16}), the propagation of turbulence is entirely in one direction, and isolated propagating disturbances are seen in the simulation, described variously as avalanches and bursts in previous work\citep{ben_bursts,candy_d3d,vanwyck}. Propagating phenomena have been frequently observed in tokamak turbulence simulations, especially when a background shear flow is imposed. Although these features are also present when no overall background flow shear is imposed\citep{ben_bursts}, we see very clear propagating features for large shears, and, as in other works\citep{vanwyck}, these become more isolated as the shearing rate approaches the critical value. The propagating edge state (fig \ref{fig:tempzonal_edge})
has some features in common with the late-time nonlinear state (figs. \ref{fig:timeevol_sp15} and \ref{fig:tempzonal_turb}), such as similar propagation velocities (to within 10$\%$) and spatial scales (a comparison of typical amplitudes can be seen in fig \ref{fig:critscan}). Both are associated with a moving turbulence front supporting a moving zonal electric field that destabilizes the system in front of the turbulence front, so can both be seen as a traveling excitation wave. Near the critical flow-shear beyond which turbulence is quenched, there structures are not particularly localised in the $y$-direction (fig \ref{fig:critstate_standard}), unlike those seen in certain more detailed simulations\citep{vanwyck}; the detailed structure of the propagating features is somewhat parameter dependent, and we present a second example of a propagating state in a lower-aspect ratio, zero magnetic shear simulation (fig. \ref{fig:critstate_lowshear}). The phenomenological commonalities between all these observations of propagating structures in gyrokinetic simulations with background shear flows are so striking that a common origin seems like the simplest explanation.  Curved and elongated eddies, associated with a self-driven radially propagating zonal shear flow appear to be a ubiquitous phenomenon in tokamak turbulence simulations. Understanding what physical and numerical parameters control the details of these structures, however, still requires further study.

\begin{figure}
\begin{subfigure}{0.5\textwidth}
\includegraphics[width=7cm]{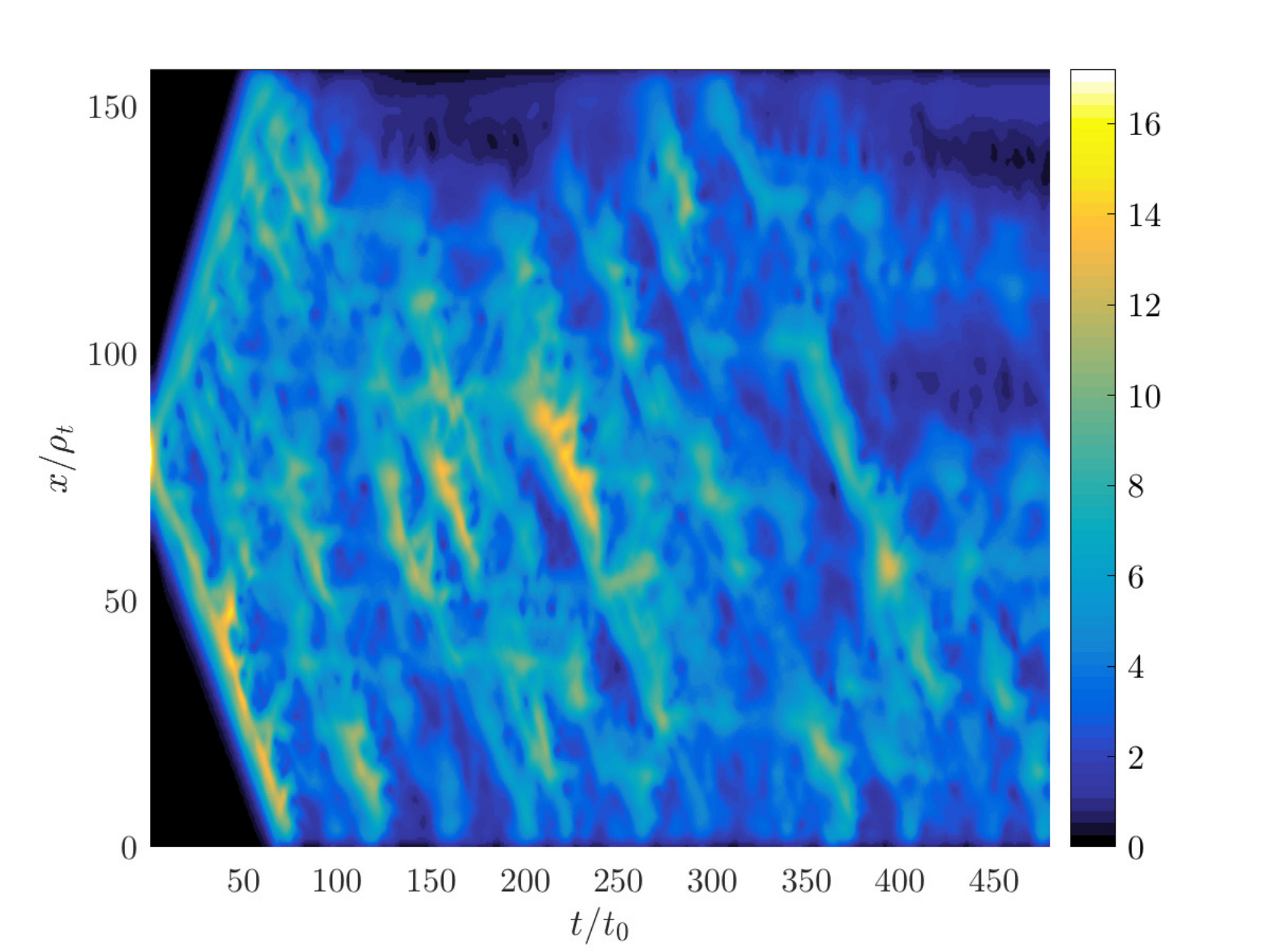} \caption{} \label{fig:burst_dirichlet_0.06}
\end{subfigure}
\begin{subfigure}{0.5\textwidth}
\includegraphics[width=7cm]{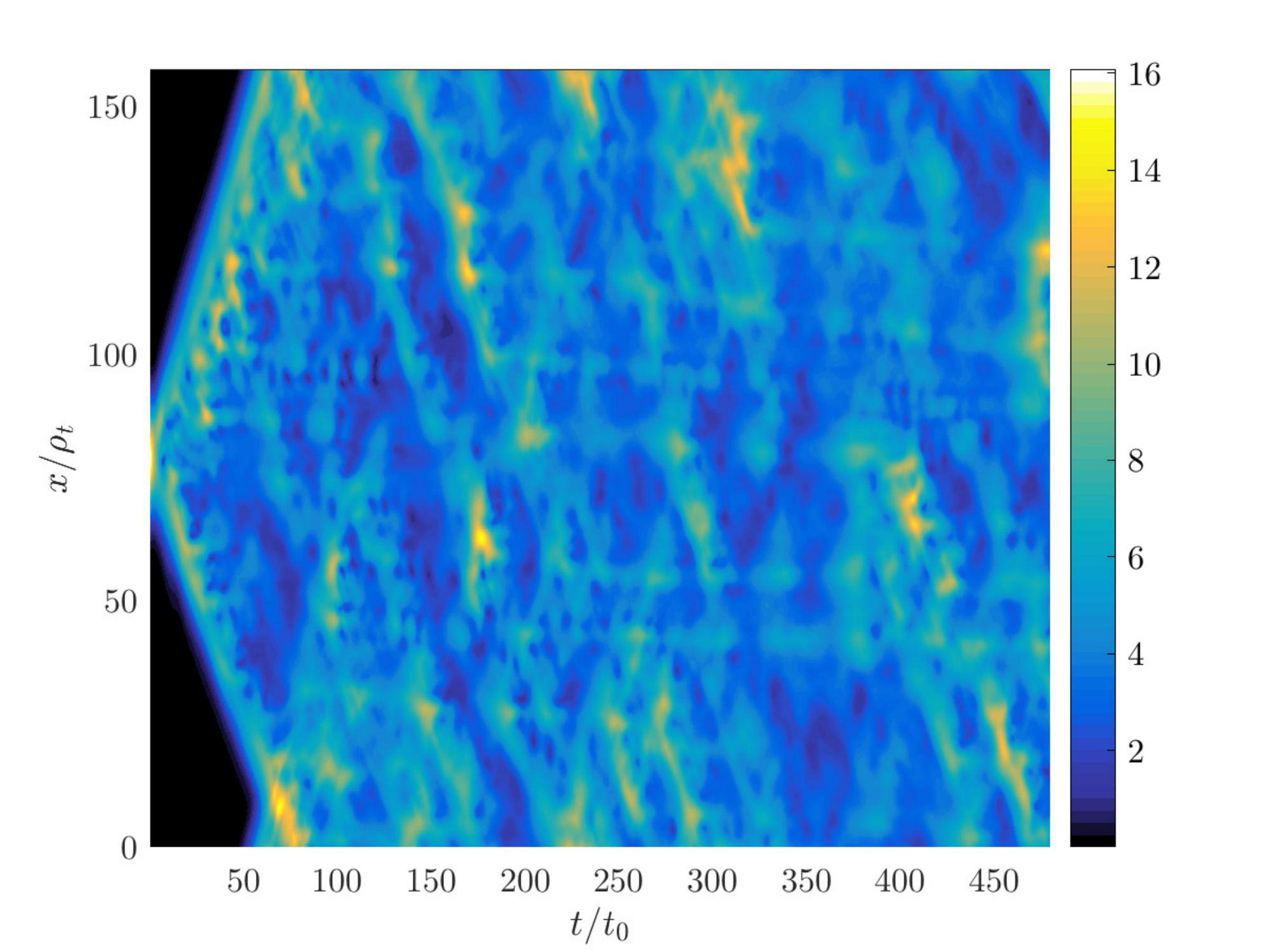} \caption{} \label{fig:burst_periodic_0.06}
\end{subfigure}

\begin{subfigure}{0.5\textwidth}
\includegraphics[width=7cm]{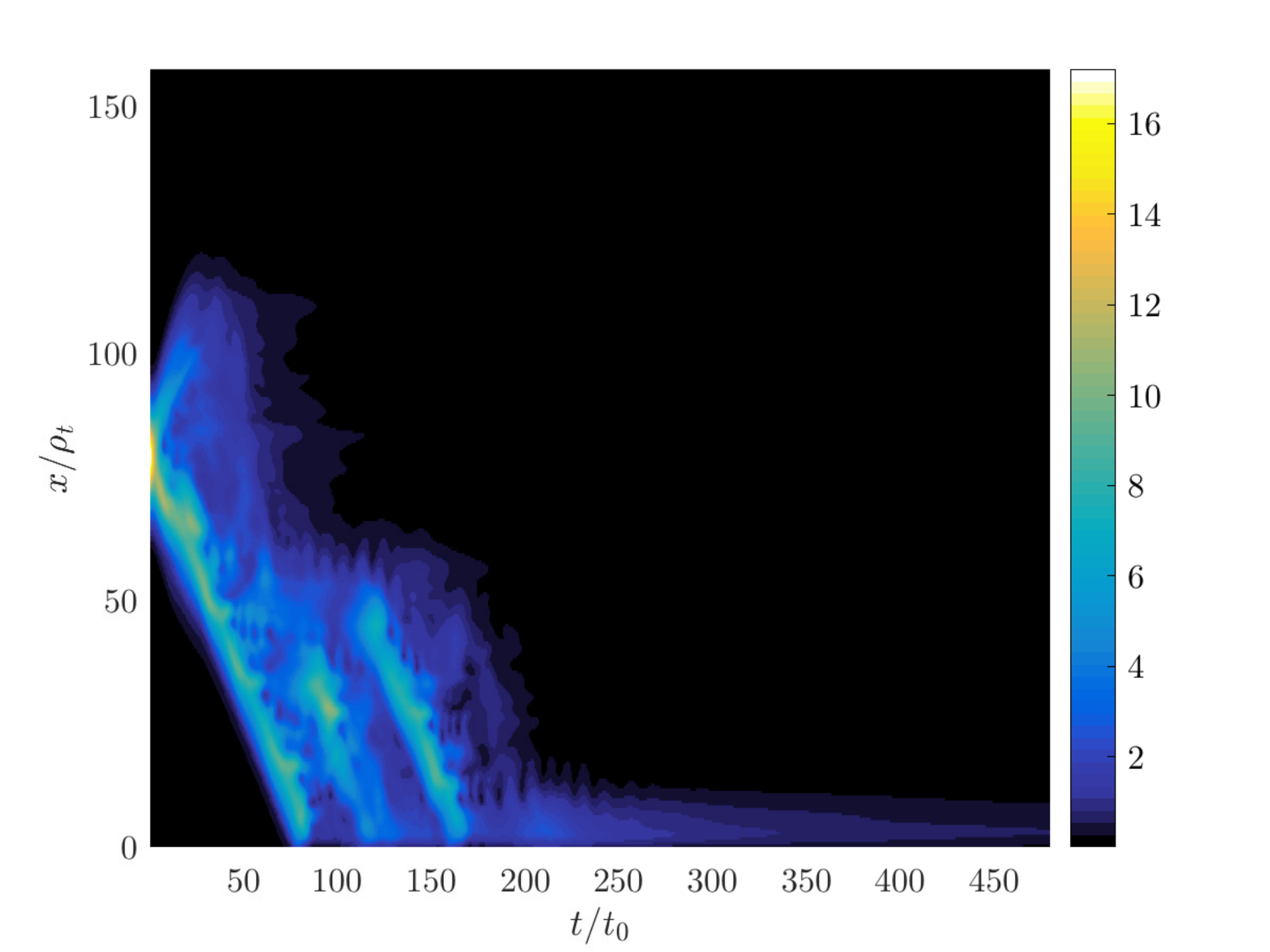} \caption{} \label{fig:burst_dirichlet}
\end{subfigure}
\begin{subfigure}{0.5\textwidth}
\includegraphics[width=7cm]{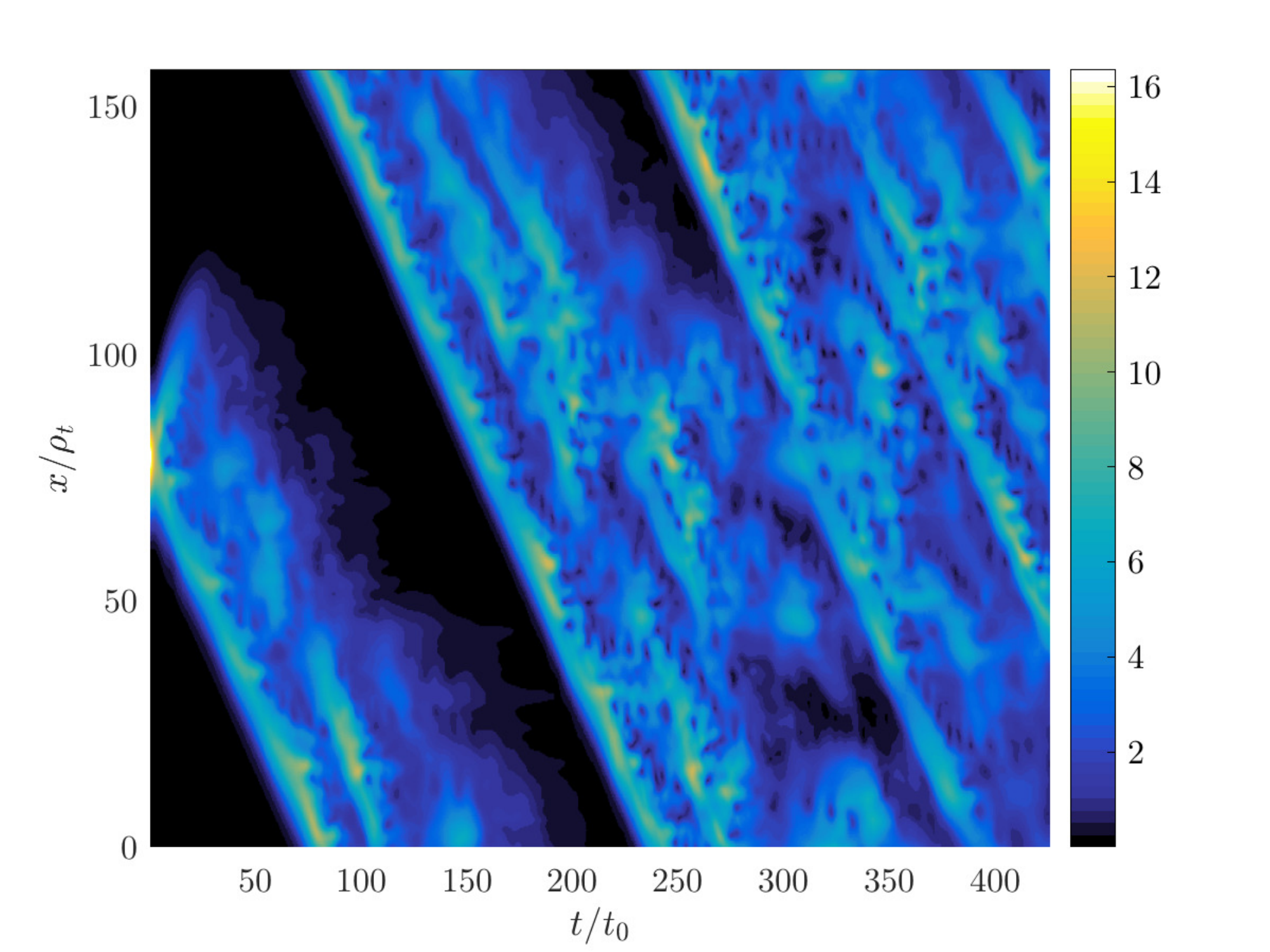} \caption{} \label{fig:burst_periodic}
\end{subfigure}

\caption{ Mean $\phi^2$ (averaged over $y$) at the midplane versus time and position in a simulation with $S = 0.06 t_0^{-1}$ with (a) Dirichlet
  and (b) periodic boundary conditions and  $S = 0.12  t_0^{-1}$ with (c) Dirichlet
  and (d) periodic boundary conditions.}
\label{fig:dirichlet_vs_periodic}
\end{figure}

\begin{figure}
\includegraphics[width=9cm]{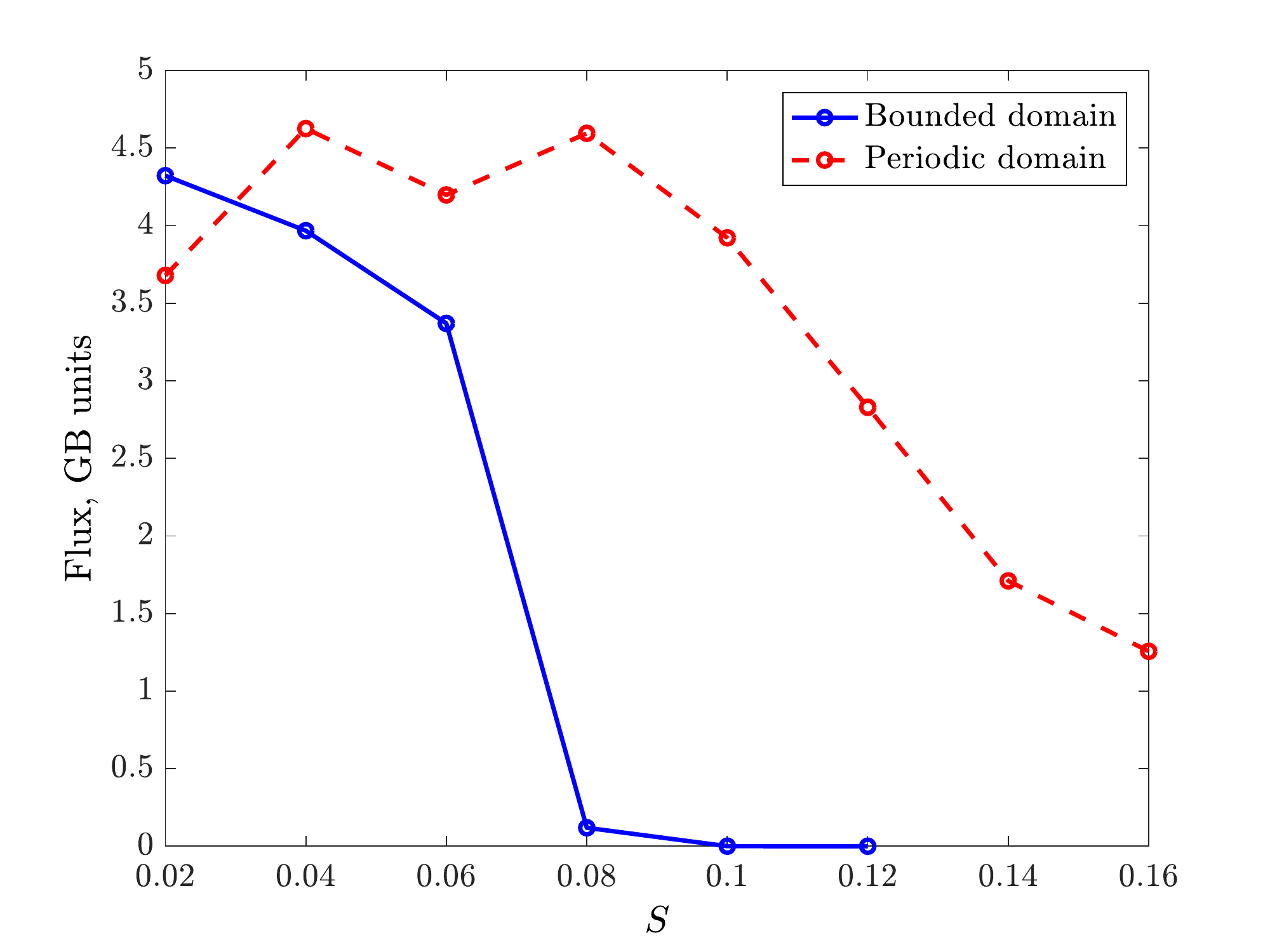}
\caption{ Volume averaged heat flux versus $S$ for Dirichlet vs periodic simulations.}
\label{fig:dirichlet_heat}
\end{figure}

As in neutral fluid simulation, if the shear is increased beyond a certain point (here, $S \sim 0.15 t_0^{-1}  $) we observe relatively long-lived turbulence that unpredictably decays to the quiescent state. It is clear from figures such as fig. \ref{fig:timeevol_sp15} that for large shear ($S \gtrsim 0.1  t_0^{-1}$), the excited region of turbulence has an overall drift, so that `puffs' of excited turbulence travel through the system, returning to a locally quiescent state after the puff has passed. Unlike in, say, pipe flow turbulence\citep{wygnanski_champagne_1973}, where these puffs travel in the direction of fluid flow, the bursts here travel is either aligned or anti-aligned with the direction of the temperature gradient.
In these simulations an unphysical periodic boundary condition is applied in the $x$ direction, so that the turbulence gradually fills the domain. We consider a simulation at $S = 0.12  t_0^{-1}$ using an `open' boundary condition (in this case applying Dirichlet conditions to $f$ and the electrostatic potential), with the standard initial condition displaced in $x$ so that it peaks at $x = 80 x_{\textrm{MAX}}$. Here, the system becomes quiescent after the puff travels to the boundary (fig. \ref{fig:burst_dirichlet} c,d). On the other hand, at lower flow shear the boundary conditions have less influence on the interior of the domain, and late-time behaviour is similar (fig. \ref{fig:burst_dirichlet} a,b). 
The sensitivity to boundary conditions is surprising in some sense because turbulent structures are very much smaller (in the $x$ direction) than the system size, and one might expect the local turbulence properties to mostly depend on local gradients, rather that the conditions at the $x$ boundaries. Nonetheless, the propagating bursts allow for patterns of activation to be set up that transmit information over longer lengthscales in the $x$ direction. We performed a scan in $S$, and find that turbulence can be sustained over a wider range of flow shear values in a periodic simulation than a bounded simulation (fig \ref{fig:dirichlet_heat}) which may explain some of the differences in earlier benchmarks\citep{Casson2011}.
The more effective quenching of the flux by background flow shear in the Dirichlet simulations does not appear to be a consequence of the specific initial conditions chosen; a simulation started with $S=0$, and restarted after turbulence has attained a near steady state with $S=0.12$ also decays to the laminar state.

\section{Scaling of the transition threshold with background flow shear}
The {\it minimal seed} is the initial simulation state of minimal amplitude that allows transition to a turbulent state in a subcritical system. The amplitude of the minimal seed may be seen as a `safety threshold' below which all perturbations will eventually decay to the laminar regime. The amplitude of the minimal seed can also be used to quantify the degree to which linear and nonlinear processes can allow amplification of small fluctuations up to turbulent levels. Examining the minimal seed amplitude compared to the edge state amplitude and the turbulence amplitude thus provides a quantification of important pieces of subcritical physics. In general the amplification factor from the minimal seed level to the edge state amplitude is not equal to the transient linear amplification factor, and in many fluids the nonlinear processes allow overall amplification factors orders of magnitude larger than linear transient growth alone.

For a general initial perturbation, the value of the transition threshold depends on the functional form of the initialization. We varied the parameters of a wavepacket-like initialization to find the `most dangerous' state with a minimal nonlinear instability threshold as an approximation to the `minimal seed'. It is in principle be possible to perform a complete minimization\citep{pringle_willis_kerswell_2012} which optimises over all possible initial states, and we were able to do this for a drift-wave model\citep{chris_bursts}. However, in the gyrokinetic context, in would require writing an adjoint gyrokinetic solver and performing subsequent computationally demanding simulations. The choice of a wavepacket type initialisation (that is, eq. \ref{eq:initialise_wavepacket}) is motivated by earlier study\citep{chris_bursts} that found this to approximately capture the true minimal seed in a drift-wave model. We performed scans of initialisation parameters at a fixed shearing rate value $0.04$ to determine the values that allowed transition at the lowest $A$ value for the standard simulation, finding
$k_{x0} \rho_0 = 0.24$,  $C_x \rho_0 = 0.1601$, $k_{y0} \rho_0 = 0.37$, $C_y \rho_0 = 0.074$ for the multi-mode simulations. In the narrow simulations other parameters are the same but we take $C_y \rightarrow 0$.

\begin{figure}
\begin{subfigure}{0.5\textwidth}
\includegraphics[width=7cm]{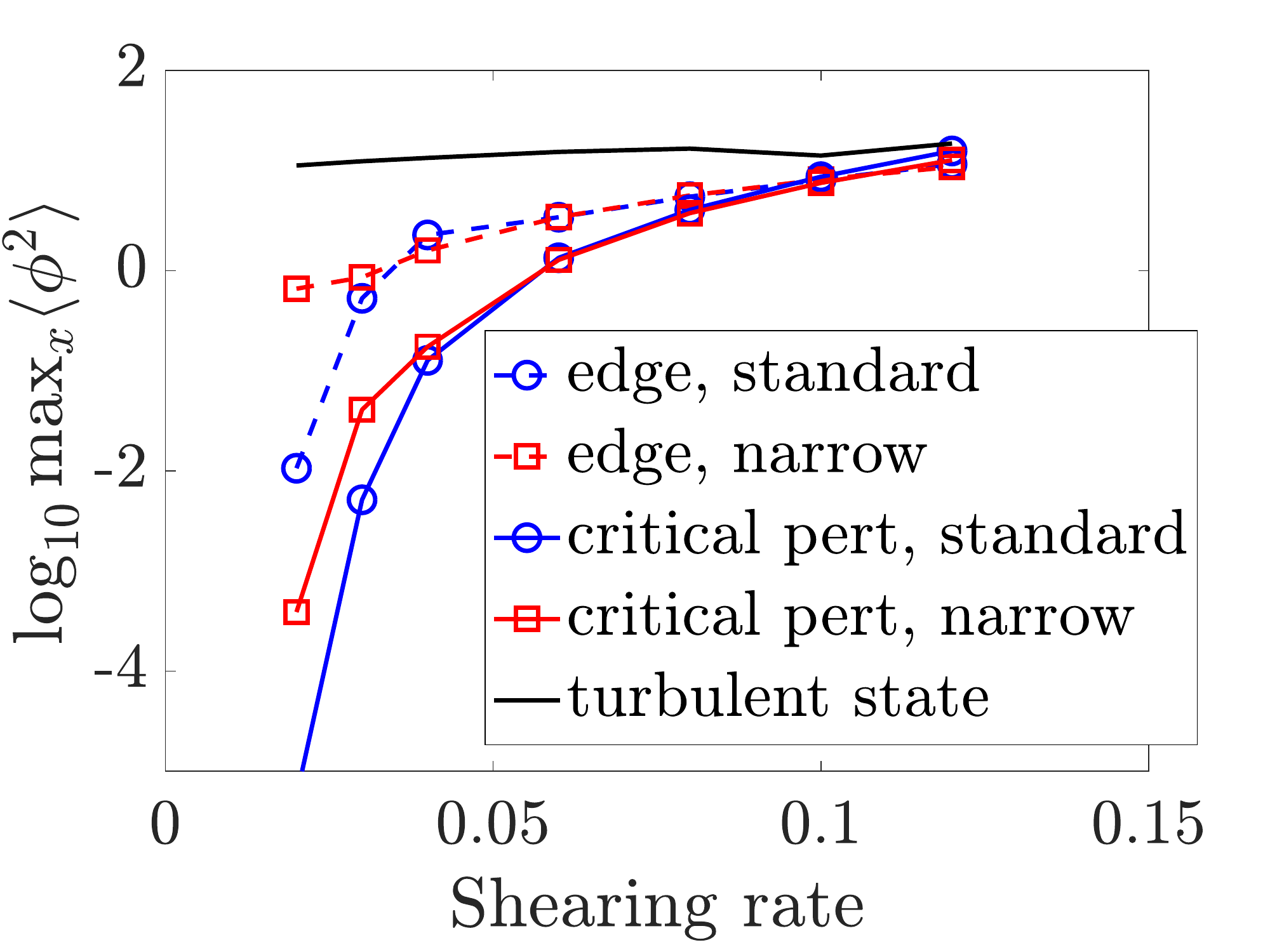} \caption{} \label{fig:supremum_phisq}
\end{subfigure}
\begin{subfigure}{0.5\textwidth}
\includegraphics[width=7cm]{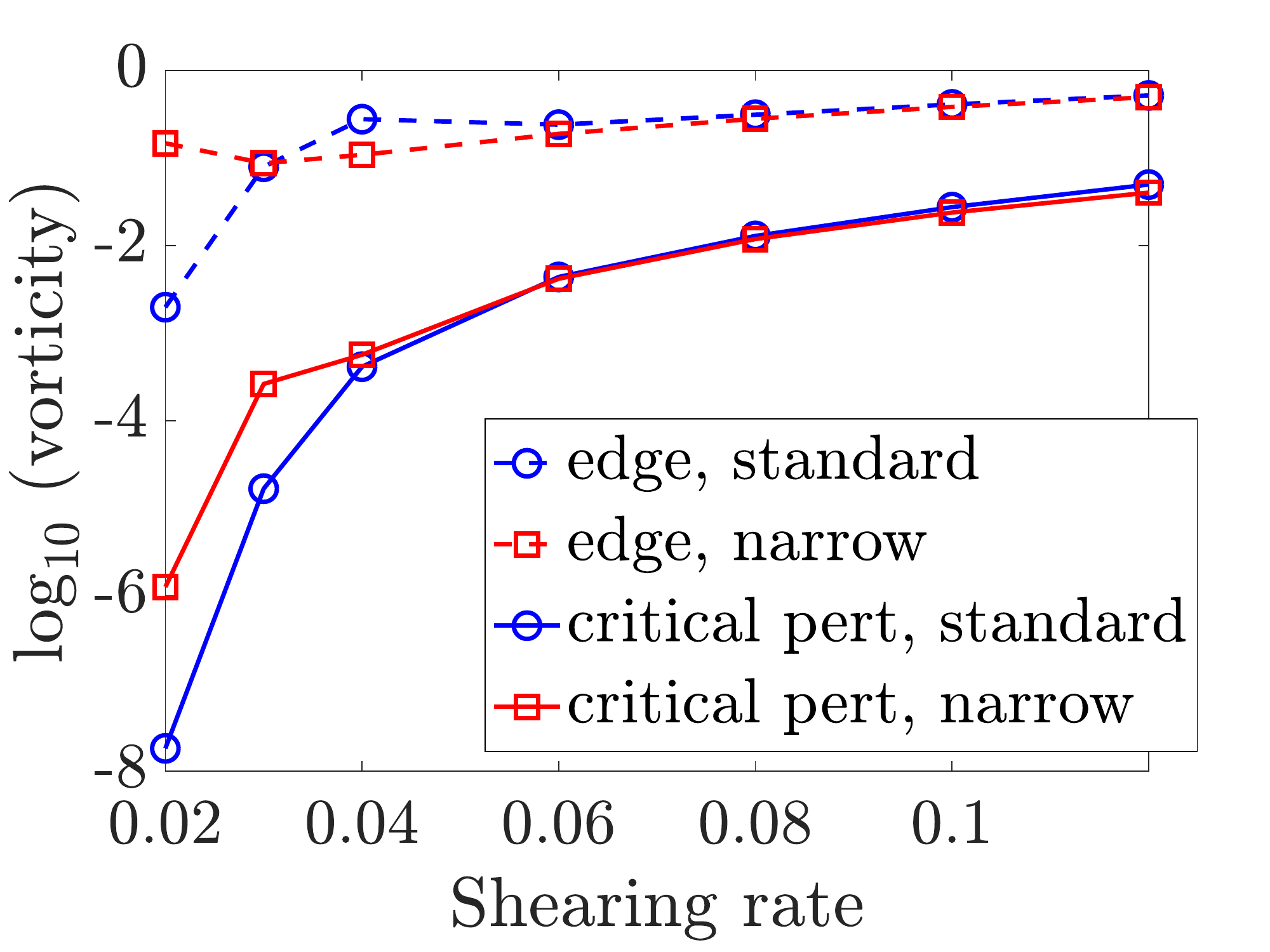} \caption{} \label{fig:logvort}
\end{subfigure}
\caption{(a) Logarithm of maximum value of squared non-zonal potential $\operatorname{max}_x \langle \phi(x,y)^2 \rangle_y$ and (b) of mean vorticity versus initial flow shear for several simulation phases, for narrow (red trace) and standard (blue trace) simulations. The amplitude of the critical state for the baseline initial perturbation shape is shown as solid traces, and the amplitude of the edge state is shown as dashed traces. At larger shearing rates $S > 0.06  t_0^{-1}$ these results for both simulation types are similar.}
\label{fig:critscan}
\end{figure}

To compare these state amplitudes here, we use two different measures. Because the minimal seed and edge states are radially localised, to compare amplitudes to the typical turbulent state, we use the maximum squared potential in $x$ (the ratio between these values should be relatively independent of the system size in $x$ unlike a spatial average measure). The other measure used is a global average vorticity. The transition threshold with shearing rate (Fig.\ref{fig:critscan}) scales roughly like $\exp(-1/S)$, except that for standard simulations at small $S$ the transition threshold drops more rapidly. Very small initial amplitudes produce instability in the small $S$ limit. The linear transient amplification also scales with $\exp(1/S)$ in these systems\citep{WaltzTrans}.

For large enough shear ($S \gtrsim 0.1 t_0^{-1}$), the transition threshold found based on the wavepacket initialisation is actually slightly higher than the edge state amplitude, when measured using the maximum measure (fig \ref{fig:supremum_phisq}) [rather than the RMS measure (fig \ref{fig:logvort})]: since the true minimum seed would have a lower transition threshold than any other state, this demonstrates that the wavepacket initialisation is not the minimum seed in this norm. This is not surprising since the reasoning used to suggest a wavepacket-like minimum seed\citep{chris_bursts} is clearly not valid except in the regime of large transient growth at low shear. Also, the amount of nonlinear transient growth depends quite strongly on which norm is used; this is expected even in linear problems, where the amount of transient amplification is a direct consequence of the choice of norm.

The edge state amplitude gives an estimate of the amplitude for which the linear and nonlinear terms balance; this reduces with small flow-shear. On the basis that the scaling of the transition threshold can be explained based on linear transient growth, the overall pathway for a near-critical mode to become unstable is hypothesised to be transient linear growth amplifying an initial perturbation, pushing it slightly beyond the edge state amplitude, after which the unstable trajectory departing from the edge state allows access to the turbulent regime. The typical situation in neutral fluid experiments, is that the transition involves several stages of linear growth chained together as a result of nonlinear effects\citep{pringle_willis_kerswell_2012}. This more complex situation appears to arise for small flow shear in the gyrokinetic simulations, where the additional toroidal modes in the standard simulation allows transition to turbulence at lower initial amplitudes (and lower edge-state amplitudes) through coupling between non-zonal modes.  The idea that scaling of subcritical thresholds in gyrokinetic systems (in that case for the maximum shearing rate at which turbulence can be sustained) can be found by considering linear transient amplification was suggested by the results of \cite{Wyk17}. This also appears to be the case in our simulations, except at low shearing rates, where the details of the nonlinear dynamics become more important (as in neutral fluids). 

A traveling-wave type edge state is found for all shear rates for the narrow simulations and for $S>0.04 t_0^{-1}$ for the standard simulations. The amplitudes of the edge state and the critical perturbation amplitude are not affected strongly by increasing the simulation box width for $S>0.06 t_0^{-1}$ where the edge-states are qualitatively similar, and the relevant nonlinearity in the critical transition to turbulence is the drift-mode/zonal-flow (and zonal gradient) interaction.


\section{Conclusions and discussions.} 
The behavior of the edge of chaos is qualitatively similar to simple plasma-interchange (PI) model\citep{chris_bursts}, strengthening the thesis\citep{ben_bursts} that qualitative features in the dynamics are the consequence of fluid-like behavior, rather than details of tokamak geometry, or subtleties in the kinetic physics. Despite the complexity of GK model compared to neutral fluid models, the edge manifold contains a quasi-traveling wave attractor, which is also seen in the PI model (note that this is attracitve only within the edge manifold, not globally). The increasing amplitude of the edge state to levels comparable to the turbulent fluctuation level (fig \ref{fig:supremum_phisq}) near the maximum background flow shear for which turbulence can be sustained, in conjunction with the relative simplicity of the edge state hints that, as in fluid turbulence theory, analysis of periodic orbits in plasma turbulence could be a powerful tool for understanding how and where turbulent states exist. The resemblances between the quasi-travelling wave in the edge of chaos, and the bursts seen in the turbulent state are notable, but as in the PI system, the nature of the relationship between these two phenomena is still unclear. We have found a way to estimate the transition threshold in this system, to quantify how robust the laminar state is against external forcings. The scaling of the transition threshold matches the PI model, except at very low shear, and the gyrokinetic system follows the same pathway to turbulence in this parameter space.

Propagating features seen in the turbulence (avalanches/bursts) have qualitative features that echo the traveling-wave edge state, with similar propagation velocities, but have stronger amplitudes and are more disordered. A simple state was seen in the limit where the flow-shear was increased to just below the threshold for sustained turbulence in \citep{Wyk17}: these investigations of the critical behaviour in these systems hint at the importance of periodic orbits in the critical dynamics of such systems. Because of the simplicity of the edge state, the mechanisms that allows the edge-state to propagate could be illustrated in detail; the traveling wave destabilizes the region in front of itself by removing the background shear flow and increasing the temperature gradient, and the tilting of the drift waves leads to a finite group velocity of the wavepacket-like finite $k_y$ modes. These propagation mechanisms appear to carry over to the avalanche/burst features in the fully turbulent state\citep{ben_bursts}. Long range propagation of features allows powerful nonlocality in these systems: at large flow shearing rate, the system is only convectively unstable, so at a fixed spatial location, the system will eventually return to a zero-flux state. On the other hand, there are a broad range of shearing rates (fig. \ref{fig:supremum_phisq}) for which a local perturbation 10\% as large as the typical turbulence level is required to initiate turbulence.

Ideas around the edge of chaos and exact solutions are well established in subcritical, neutral fluid problems. Some progress has also been made in applying them in astrophysical plasmas\citep{rincon}. Here we have taken the first steps in applying them to study tokamaks. The results show that these methods can reveal intriguing aspects of this problem, but pose as many questions as they answer. Is the edge always dominated by a simple quasi-travelling wave for all parameter regimes? Do other such states exist? Can these states be extended into the turbulent attractor? How densely is state space packed with such solutions, and how are they connected?

The quasi-travelling wave presented could only be isolated as it was linearly stable within the edge. Even with this advantage, the bisection technique required is time consuming. To pursue these problems further more advanced techniques are required. Such techniques (primarily matrix-free Newton-Krylov solvers) have been widely applied in classical fluids to find, track and analyse steady states, travelling waves and other, more complex, classes of exact solution. Implementing these techniques within existing plasma codes is an ambitious but feasible problem which this paper motivates and begins to open the door to. 



\begin{acknowledgments}
{{\em Acknowledgments.---} 
B. McMillan is partially supported by EPSRC grant no. EP/N035178/1.
C. Pringle is partially supported by EPSRC grant No. EP/P021352/1.
B. Teaca is partially supported by EPSRC grant No. EP/P02064X/1.
Simulations were performed with the support of Eurofusion and MARCONI-Fusion. We acknowledge the authors of the GKW code for developing and distributing this software tool.}
\end{acknowledgments}

\bibliographystyle{jpp}

\end{document}